\documentclass[fleqn,usenatbib]{mnras}

\usepackage{newtxtext,newtxmath}
\usepackage{soul}
\usepackage{threeparttable}

\usepackage[T1]{fontenc}

\DeclareRobustCommand{\VAN}[3]{#2}
\let\VANthebibliography\thebibliography
\def\thebibliography{\DeclareRobustCommand{\VAN}[3]{##3}\VANthebibliography}

\usepackage{graphicx}	
\usepackage{amsmath}	
\usepackage{threeparttable}
\usepackage{color}
\usepackage{xcolor}
\usepackage{textcomp}

\newcommand\lsim{\mathrel{\rlap{\lower4pt\hbox{\hskip1pt$\sim$}}
    \raise1pt\hbox{$<$}}}
\newcommand\gsim{\mathrel{\rlap{\lower4pt\hbox{\hskip1pt$\sim$}}
    \raise1pt\hbox{$>$}}}

\title[Gas-assisted formation of
binaries]{Binary formation through gas-assisted capture and the implications for stellar, planetary and compact-object evolution}

\author[Rozner, Generozov \& Perets]{
Mor Rozner,$^{1}$\thanks{E-mail: morozner@campus.technion.ac.il}, 
Aleksey Generozov$^{1}$ and 
Hagai B. Perets$^{1}$
\\
$^{1}$Technion - Israel Institute of Technology, Haifa, Israel, 3200003\\
}

\date{Accepted XXX. Received YYY; in original form ZZZ}

\pubyear{2022}

\begin{document}
\label{firstpage}
\pagerange{\pageref{firstpage}--\pageref{lastpage}}
\maketitle

\begin{abstract}
Binary systems are ubiquitous and their formation requires two-body interaction and dissipation. In gaseous media, interactions between two initially unbound objects could result in gas-assisted binary formation, induced by a loss of kinetic energy to the ambient gas medium. Here we derive analytically the criteria for gas-assisted binary capture through gas dynamical friction dissipation. We validate them with few-body simulations and explore this process in different gas-rich environments: gas-embedded star-forming regions (SFR), gas-enriched globular clusters, AGN disks and protoplanetary-disks. We find that gas-assisted binary capture is highly efficient in SFRs, potentially providing a main channel for the formation of binaries. It could also operate under certain conditions in gas-enriched globular clusters. Thin AGN disks could also provide a fertile ground for gas-assisted binary capture and in particular the formation of black-hole/other compact object binaries, the production of gravitational-wave (GW) and other high-energy transients. Large-scale gaseous disks might be too thick to enable gas-assisted binary capture and previous estimates of the production of GW-sources could be overestimated, and sensitive to specific conditions and the structure of the disks. In protoplanetary-disks, while gas-assisted binary capture can produce binary KBOs, dynamical friction by small planetsimals is likely to be more efficient. Overall, we show that gas-assisted binary formation is robust and can contribute significantly to the binary formation rate in many environments. In fact, the gas-assisted binary capture rates are sufficiently high such that they will lead to multicaptures, and the formation of higher multiplicity systems. 
\end{abstract}




\section{Introduction}

Binary systems are ubiquitous over a wide range of scales and in different astrophysical systems, from binary planetesimals in the Solar system, through stellar binaries and compact objects and up to the scales of binary massive black holes.  Indeed, the majority of stars reside in binaries, or even higher multiplicity systems (e.g. \citealp{Raghavan2010,Sana2012,DucheneKraus2013,MoeDiStefano2017}), and a large fraction of KBOs (Kuiper-belt objects) reside in binaries (see a detailed review in \citealp{Noll2008}). 

Binaries play a key role in the dynamics and evolution of stars and compact objects. In particular, close interactions between binary companions could lead to mass-transfer or even mergers and collisions of the binary components. In turn, these interactions could give rise to the production of exotic stars and compact objects, which could otherwise not form from the evolution of single stars. Some compact binaries produce high energy emission (e.g. X-ray binaries) and mergers could result in explosive transient events such as supernovae (SNe), gamma-ray bursts (GRBs) and/or the production of gravitational-wave (GW) sources. 

Understanding the formation of binaries and their properties is therefore essential for decoding the evolution of stellar and planetary systems.

Several prominent binary formation channels were explored in the literature (see \citealt{LeeOffner+2020} and the references therein). These could generally be divided between primordial formation of binaries, where the binary components form together as a bound systems, and dynamical formation channels, where each of the binary component forms independently, and later dissipative processes bind them together to form a binary. The former involves the fragmentation of a bound blob of gas/dust in which two objects form and orbit each  other (e.g. \citealp{Nesvorny2010}). The latter involves dissipation mechanisms, where various channels were suggested to form binaries: (1) Tidal forces \citep{Fabian1975,PressTeukolsy1977}, gravitational wave (GW) emission, or even collisions, all of which become effective only through very close encounters between the progenitor unbound components. (2) Three-body encounters, where the gravitational perturbation transfers kinetic energy between the components, until one is ejected with higher velocity, leaving behind a bound binary \citep{AarsethHeggie1976}. (3) Dynamical friction \citep{GoldreichSari2002} and gas dynamical friction \citep{Tagawa2020} where two objects embedded in a sea of far less massive particles or in gas dissipate their excess kinetic energy to the ambient medium, leaving behind a bound binary. Although gas-rich environments are quite common, the latter gas-assisted capture scenario was little studied, although recently these environments gained more focus in this context (e.g. \citealp{LiLai2022,Rowan22,BoekholyKocsis2022}). Here we explore this scenario analytically and using few-body simulations. We provide the specific conditions in which gas-assisted captures occur, and the dependence on  the progenitor components, the ambient gas environment and the limitations put by external potentials. We then use these calculations to explore the implications of gas-assisted capture to the formation of binaries over a wide range of environments, assess its importance and the capture rates expected from this channel. In particular, we focus on AGN disks, where gas-assisted capture could play a key role in the production of GW sources; in star-forming regions, where gas-assisted capture could serve as the main channel for binary formation;  and in globular clusters and protoplanetary disk, for which we find the gas-assisted capture is likely to be far less efficient. 

 In \S~\ref{sec:GDF intro} we introduce the model of gas dynamical friction. In \S~\ref{sec:gas assisted capture} we discuss the conditions for gas-assisted binary formation and derive criteria for such a capture, using numerical and analytical methods. We then discuss the implications of our results and the probability for gas-assisted captures in several astrophysical environments (\S~\ref{sec:discussion}): SF environments (\S~\ref{subsub:SF}), second or later generations of GCs (\S~\ref{subsub:GC}), AGN disks (\S~\ref{subsub:AGN}) and the Kuiper-belt (\S~\ref{subsub:PPD}). In \S~\ref{sec: caveats}, we discuss the caveats of our model.
In § \ref{subsec:heating}, we discuss the heating and cooling related to gas-assisted captures.
 Finally, in \S~\ref{sec:summary}, we summarize our findings.

\section{Gas Dynamical Friction}\label{sec:GDF intro}

 There are several models to describe the dynamics of objects in gas, among them are evolution in gas-rich minidisks (e.g. \citealp{Stone2017}) and gas dynamical friction. Here, unless stated otherwise, we will focus on gas dynamical friction (GDF). 
 
The GDF force on an object with mass $m$ is \citep{Ostriker1999},

\begin{align}\label{eq:FGDF}
\textbf{F}_{\rm GDF}=-\frac{4\pi G^2m^2 \rho_g}{v_{\rm rel}^3}\textbf{v}_{\rm rel}
I(v/c_s)
\end{align}

\noindent
where $G$ is the gravitational constant, $\rho_g$ is the gas density, $c_s$ is the sound speed, and $\mathbf{v_{\rm rel}}$ is the relative velocity between the object and the gas. The function $I$ is given by 

\begin{align}
I(\mathcal M)=
    \begin{cases}
        \frac{1}{2} \log(1-\mathcal M^{-2})+\ln \Lambda, & \mathcal M> 1\\
        \frac{1}{2} \log\left(\frac{1+\mathcal M}{1-\mathcal M}\right)-\mathcal M, & \mathcal M<1\\
    \end{cases}
    \label{eq:I}
\end{align}
Here, $\ln \Lambda$ is the Coulomb logarithm.\footnote{For a finite time perturbation $\ln \Lambda$ is a function of time.} For $\mathcal M\gg 1$, $I$ is nearly independent of the Mach number. Thus, for simplicity, we use the following modified function instead

\begin{align}
    I(\mathcal M)=
    \begin{cases}
        \ln\Lambda & \mathcal M\geq 1\\
        \min\left\{\ln \Lambda, \frac{1}{2}\log\left(\frac{1+\mathcal M}{1-\mathcal M}\right)-\mathcal M\right\} & \mathcal M<1
    \end{cases}
    \label{eq:Ib}
\end{align}
Following \citep{Tagawa2020}, we assume $\ln \Lambda=3.1$. For numerical stability, we use the series expansion $I(\mathcal M)\approx \mathcal M^3/3+\mathcal M^5/5$ for M<0.02 in our numerical calculations.
The energy and angular momentum of the captured binary are (correspondingly)

\begin{align}
E=-\frac{Gm_1m_2}{2a}, \ 
L=\mu_{\rm bin}\sqrt{GM_{\rm bin}a(1-e^2)},
\end{align}
respectively.

\section{Gas-assisted capture}\label{sec:gas assisted capture}

Energy dissipation induced by GDF could lead, under conditions we describe later, to binary formation, similarly to the L2 mechanism \citep{GoldreichSari2002}.
While L2 relies on dissipation induced by dynamical friction by other Kuiper-belt objects, we focus on GDF (see also \citealp{Tagawa2020}).

Generally, capture occurs if the energy dissipated during the passage of the objects is larger than the initial free unbound energy. Then the binary will be left bound at least momentarily. However, further evolution could unbind the binary or harden it. The discussion on further evolution is left for future work. 


In this section we derive both analytically and numerically the conditions for gas-assisted binary capture.

\subsection{Threshold velocity for capture}\label{subsec:threshold velocit}

The maximum initial velocity where capture occurs can be estimated by equating the work done by dynamical friction to the initial energy of the (unbound) orbit. We also require that capture occurs within the Hill sphere, where the gravity of the two-body system dominates tidal forces. If the separation of particles exceeds the Hill radius. tidal forces from other objects (e.g. the central black hole or star), would dominate the gravity of the two bodies, and they would be torn apart. 

Thus capture occurs if,
\begin{align}
\small
\frac{1}{2}\mu v_\infty^2=\Delta E_{\rm GDF}\approx \textbf F_{\rm GDF}(m_1,v_1,v_g)\cdot\boldsymbol\ell_1+\textbf F_{\rm GDF}(m_2,v_2,v_g)\cdot\boldsymbol\ell_2
\end{align}

\noindent
where $\ell_i$ is the typical length scale in which mass, $m_i$, dissipates its energy; $\mu$ is the reduced mass of the two-body system; and $v_{\infty}=v_1-v_2$ is the relative velocity at infinity. 

For simplicity, we assume that the gas center-of-mass is at rest with respect to the binary center-of-mass, we discuss the effect of a headwind in \S~\ref{subsec:headwind}. Thus, $v_1$ and $v_2$ are the initial velocities in the center-of-mass frame and $m_1 v_1=m_2 v_2$.  It should be noted that by construction, the momentum could not be conserved, due to the action of the external force, but we consider only a local conservation as an approximation for short timescales. 
In general, $\ell_1/\ell_2=q^\alpha$. The power-law index, $\alpha$, is $2$ in the subsonic case and $5$ in the supersonic case (see Appendix~\ref{sec:vel}). The maximum of $\ell_1$ and $\ell_2$ cannot exceed the Hill radius, to ensure a gravitational interaction between the objects.

\begin{table}
	\caption{\label{table:Critical velocities, vg=0} Maximum velocities for capture in different regimes of GDF, with no headwind.}
    \begin{threeparttable}
	\begin{tabular*}{\columnwidth}{@{\extracolsep{\fill}}lcc}
		  &  Supersonic & Subsonic\\
		  \hline
		 \hline
		 Unfocused
		 & $v_x q^{1/4} (1+q)^{3/4}$ & $v_s q$
		 \\
		 \hline
		 Focused & $\frac{v_x^2}{v_{\rm esc}} \frac{(1+q)^{1/2}}{q}$ & $\frac{\sqrt{8qv_s v_{\rm esc}  }}{1+q}$\\
    \hline \hline
	\end{tabular*}
    \begin{tablenotes}
    \item $v_x=\left(8 \pi G^2 \rho_g m_{\rm bin} R_{\rm Hill} \ln \Lambda \right)^{1/4}$
    \item $v_{\rm esc}=\sqrt{\frac{2 G m_{\rm bin}}{R_{\rm Hill}}}$
    \item $v_s=\frac{8 \pi G^2 \rho_g m_{\rm bin} R_{\rm Hill}}{3 c_s^3}$
    \end{tablenotes}
	\end{threeparttable}
\end{table}

In the supersonic regime, the timescale for deceleration decreases rapidly with the stars' velocity. Thus, the timescale for the binary elements to evolve after capture will be shorter than the initial capture timescale. By geometry, the binary's initial period will be comparable to or greater than this capture time. Thus, the binary's orbital elements will evolve over a dynamical timescale (i.e. they will change significantly over one orbital period). 

We present a derivation of the threshold velocity for capture in Appendix~\ref{sec:Threshold Derivation}, and summarize the results below.
In the supersonic regime, capture occurs if the initial relative velocity at infinity is less than
\begin{align}\label{eq:supersonic thres}
    &v_{c,1}=
    \begin{cases}
        v_x q^{1/4} \left(1+q\right)^{3/4}, & v_x\gg v_{\rm esc}\\
        \frac{v_x^2}{v_{\rm esc}} \frac{(1+q)^{1/2}}{q}, & v_x\ll v_{\rm esc}
    \end{cases}\nonumber\\
    &v_x=\left(8 \pi G^2 \rho_g m_{\rm bin} R_{\rm Hill} \ln \Lambda \right)^{1/4}\nonumber\\
    &v_{\rm esc}=\sqrt{\frac{2 G m_{\rm bin}}{R_{\rm Hill}}}
\end{align}
where $R_{\rm Hill}$ is the Hill radius, $m_{\rm bin}$ is the total mass, and $q$ is mass ratio (between the secondary and primary masses). The first line corresponds to the threshold, neglecting the effects of gravitational focusing. This estimate is appropriate if the relative velocity at infinity is much greater than the escape speed at the Hill radius.  Conversely, the second line corresponds to the threshold, assuming gravitational focusing is dominant (i.e. we approximate the particle trajectories as parabolic in estimating the work done as the particles cross the Hill sphere). This is appropriate if the relative velocity at infinity is much less than the escape speed at the Hill radius. In the supersonic regime, we use the unfocused estimate if $v_x>v_{\rm esc} (1+q)^{1/4} q^{5/4}$, where the two estimates are the same. Otherwise, we use the focused estimate. We refer to these two cases as the ``unfocused regime’’ and the ``focused regime.''

In the subsonic regime capture occurs as long as the velocity at infinity is less than 
\begin{align}
        &v_{c,2}=
    \begin{cases}
        v_s q, & v_s\gg v_{\rm esc}\\
        \sqrt{v_s v_{\rm esc} \frac{8 q}{(1+q)^2}}, & v_s\ll v_{\rm esc}
    \end{cases}\label{eq:subsonic thresh}\nonumber\\
    &v_s=\frac{8 \pi G^2 \rho_g m_{\rm bin} R_{\rm Hill}}{3 c_s^3}
\end{align}

\noindent
Once again we assume the transition between the focused and unfocused regimes occurs where the two estimates for the threshold become equal (when $v_s=\frac{8 v_{\rm esc}}{q (1+q)^2}$).

The thresholds in equations~\ref{eq:supersonic thres} and~\ref{eq:subsonic thresh} are approximate, since (in the unfocused regime) we use the initial velocity to estimate the energy dissipated. This is justified, because most energy will be dissipated at large velocities, due to the quadratic dependence of kinetic energy on velocity. For given gas properties and masses, one does not know a priori whether the threshold velocity for capture will be subsonic or supersonic (and whether to use the estimate in equation~\ref{eq:supersonic thres} or~\ref{eq:subsonic thresh}). Generally, only one of equation~\eqref{eq:supersonic thres} and equation~\eqref{eq:subsonic thresh} will give a consistent result. If $v_{c,1} \lsim v_{c,2}$ then $\frac{v_c}{c_s} \gsim 1$. Both estimates of the threshold are supersonic, and thus the supersonic estimate ($v_{c,1}$) should be used. Conversely, if $v_{c,1} \gsim v_{c,2}$ then $\frac{v_c}{c_s} \lsim 1$. Both estimates are subsonic, and thus the subsonic estimate ($v_{c,2}$). In general, the threshold for capture can be estimated using 

\begin{equation}
    v_c={\rm min}\{v_{\rm {c,1}}, v_{\rm{c,2}}\}.
    \label{eq:capt}
\end{equation}

\noindent
We validate equation~\eqref{eq:capt} with numerical simulations in \S~\ref{sec:numerical validation}.

\begin{figure}
    \includegraphics[width=\columnwidth]{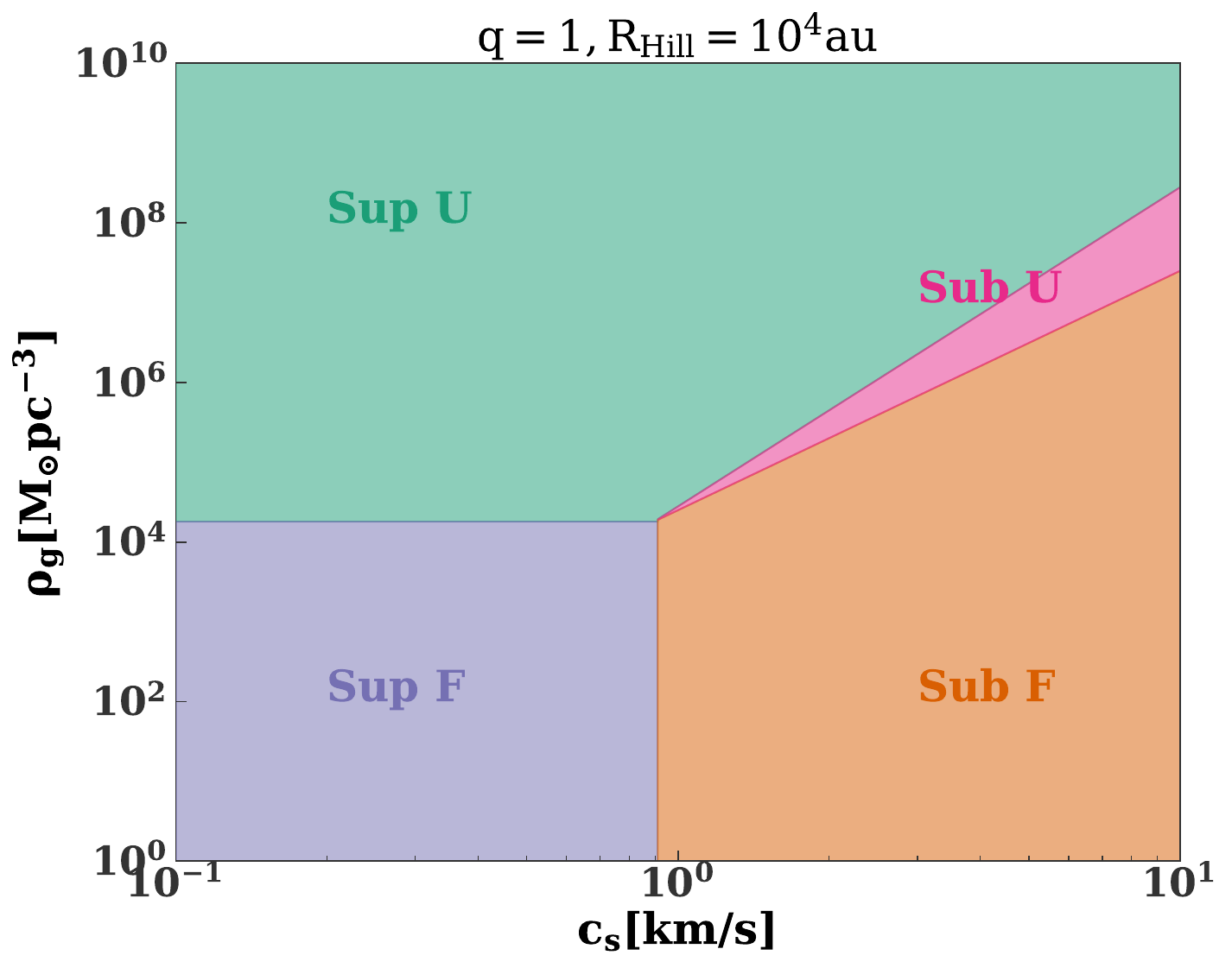}
    \includegraphics[width=\columnwidth]{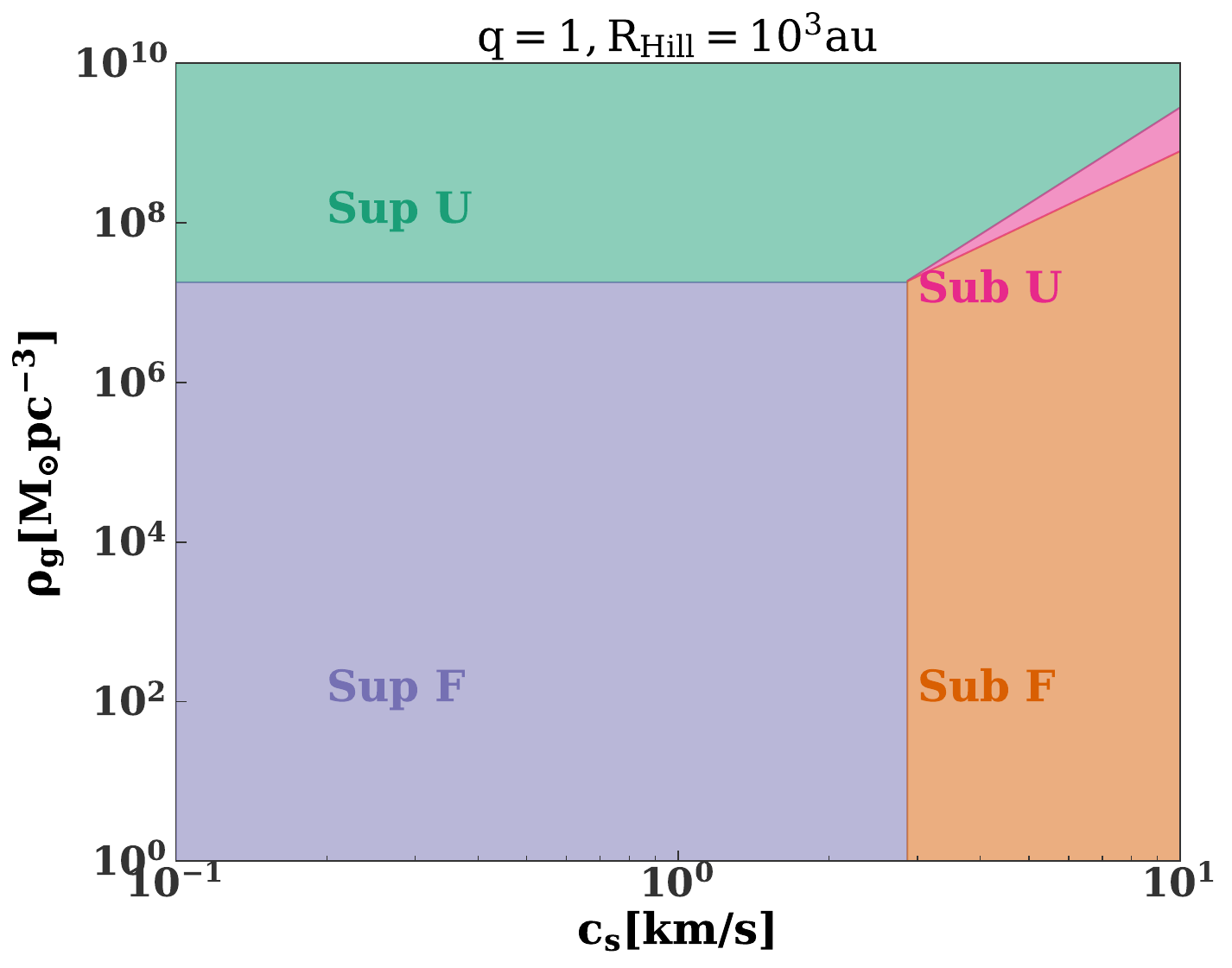}
    \includegraphics[width=\columnwidth]{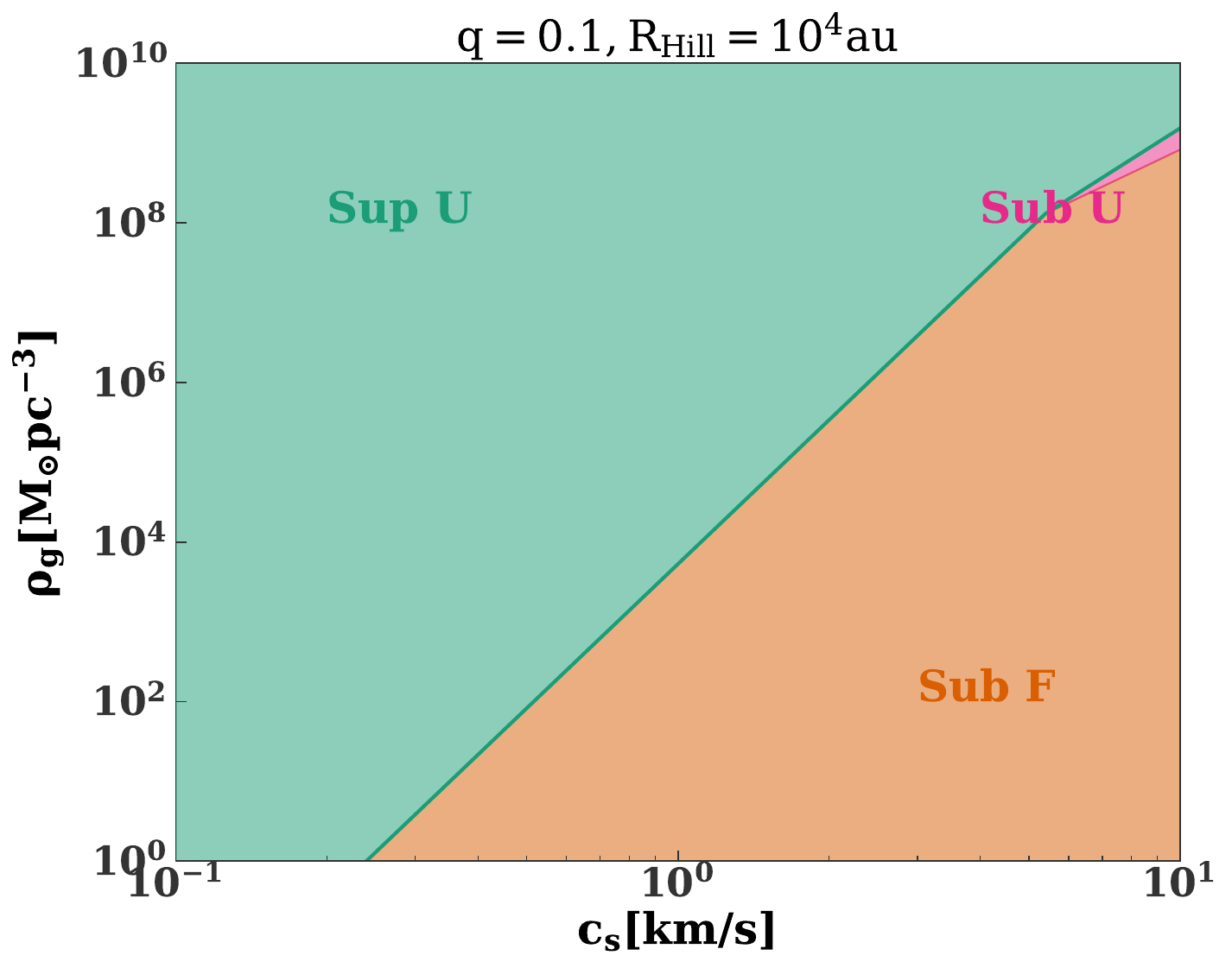}
    \caption{Separation between different capture regimes (see the text for details) as a function of gas density and sound speed for two $10 M_{\odot}$ stars  or compact objects. ``Sup'' and ``Sub'' correspond to supersonic and subsonic, respectively, while ``F'' and ``U'' correspond to focused and unfocused.}
    \label{fig:Capture regimes}
\end{figure}

Table~\ref{table:Critical velocities, vg=0} summarizes the threshold capture velocities in different regimes. Figure~\ref{fig:Capture regimes} shows delineation between regimes as a function of the gas density and sound speed.  

\subsection{Numerical validation}
\label{sec:numerical validation}

In order to validate our analytic results, we make use of few-body numerical simulations with an added GDF force.
We place two particles on an initially unbound orbit, and numerically integrate them forward in time under the influence of GDF. (See equation~\ref{eq:FGDF} and the discussion there.) We assume the initial separation corresponds to the Hill radius, although we did not explicitly include a tidal field in our simulations. For simplicity, the components of the separation parallel and perpendicular to the direction of motion are the same. In other words, the initial relative velocity is misaligned by 45\textdegree\,  with respect to initial separation.\footnote{The threshold velocity has a weak dependence on the impact parameter, $b$, viz. $v_c\propto \left({1-\frac{b^2}{R_{\rm Hill}^2}}\right)^\xi$, where $\xi$ is 1/8 (1/2) in the supersonic (subsonic) regime. We neglect this correction.}

We evolve the two stars with the IAS15 integrator \citep{rein.spiegel2015} in \texttt{REBOUND} \citep{rein.liu2012}. GDF is included via \texttt{REBOUNDx} \citep{tamayo+2019}. The gas medium has a constant density (in space and time) and is at rest with respect to the binary center-of-mass initially.

\begin{figure}
    \includegraphics[width=\columnwidth]{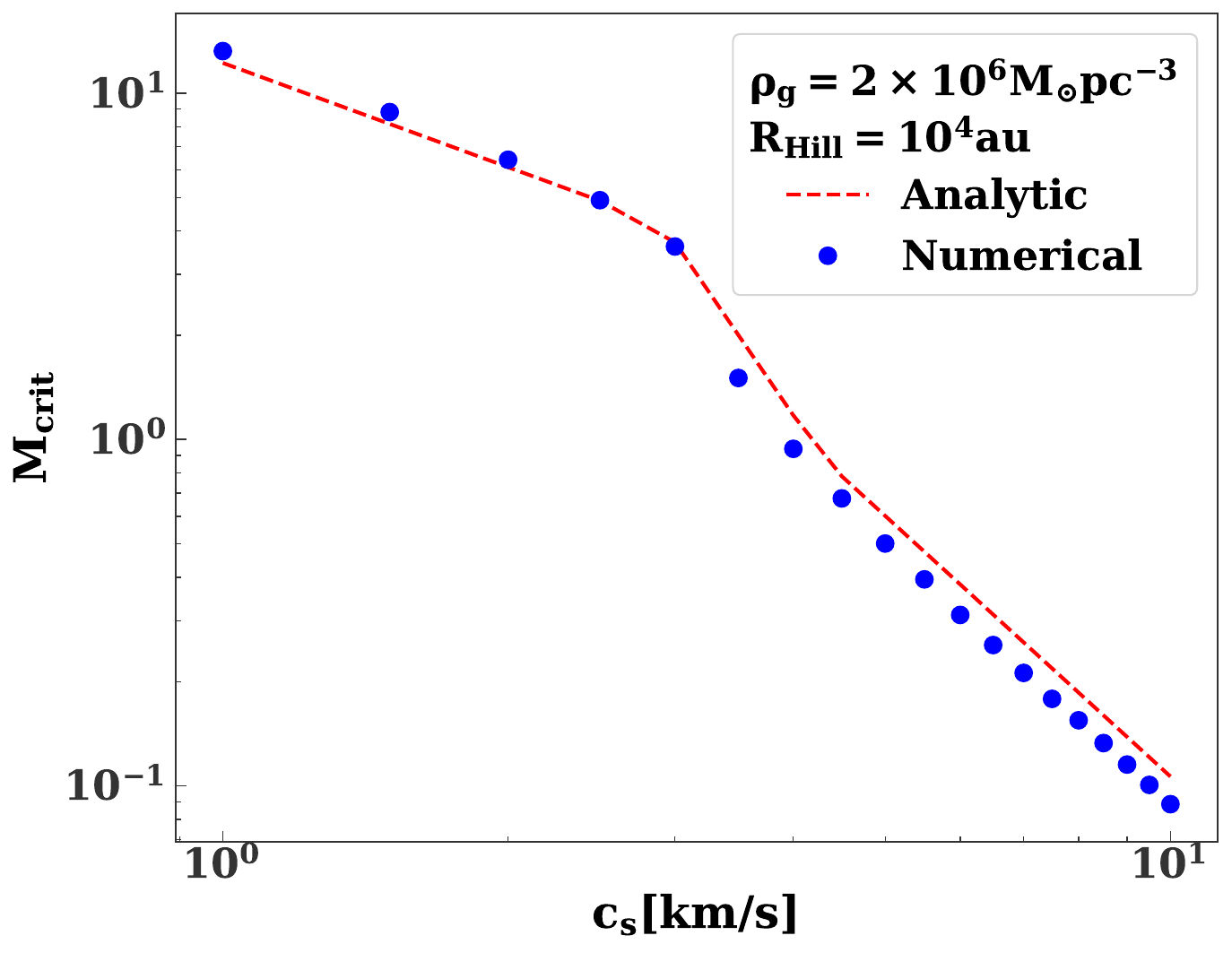}
    \includegraphics[width=\columnwidth]{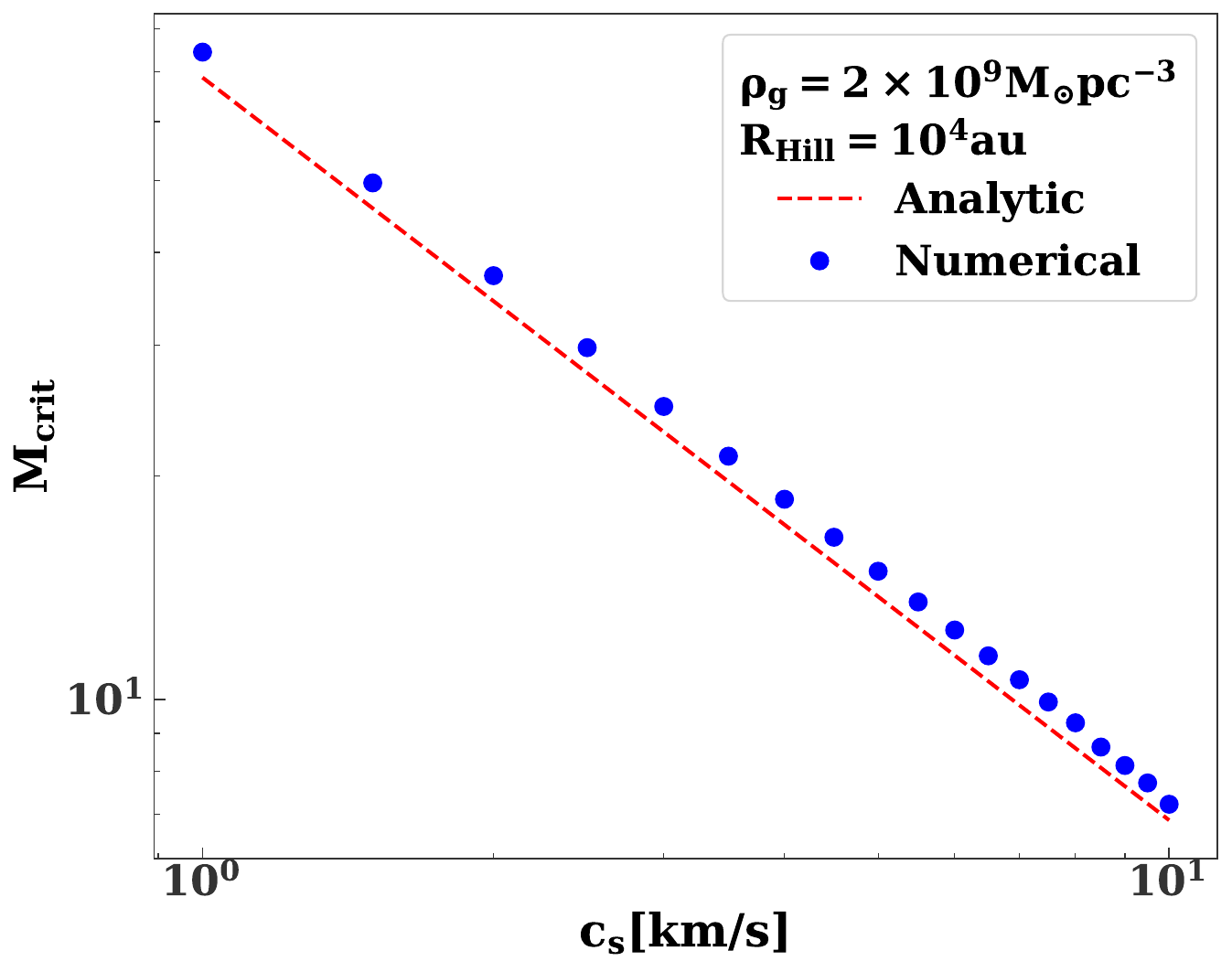}
    \includegraphics[width=\columnwidth]{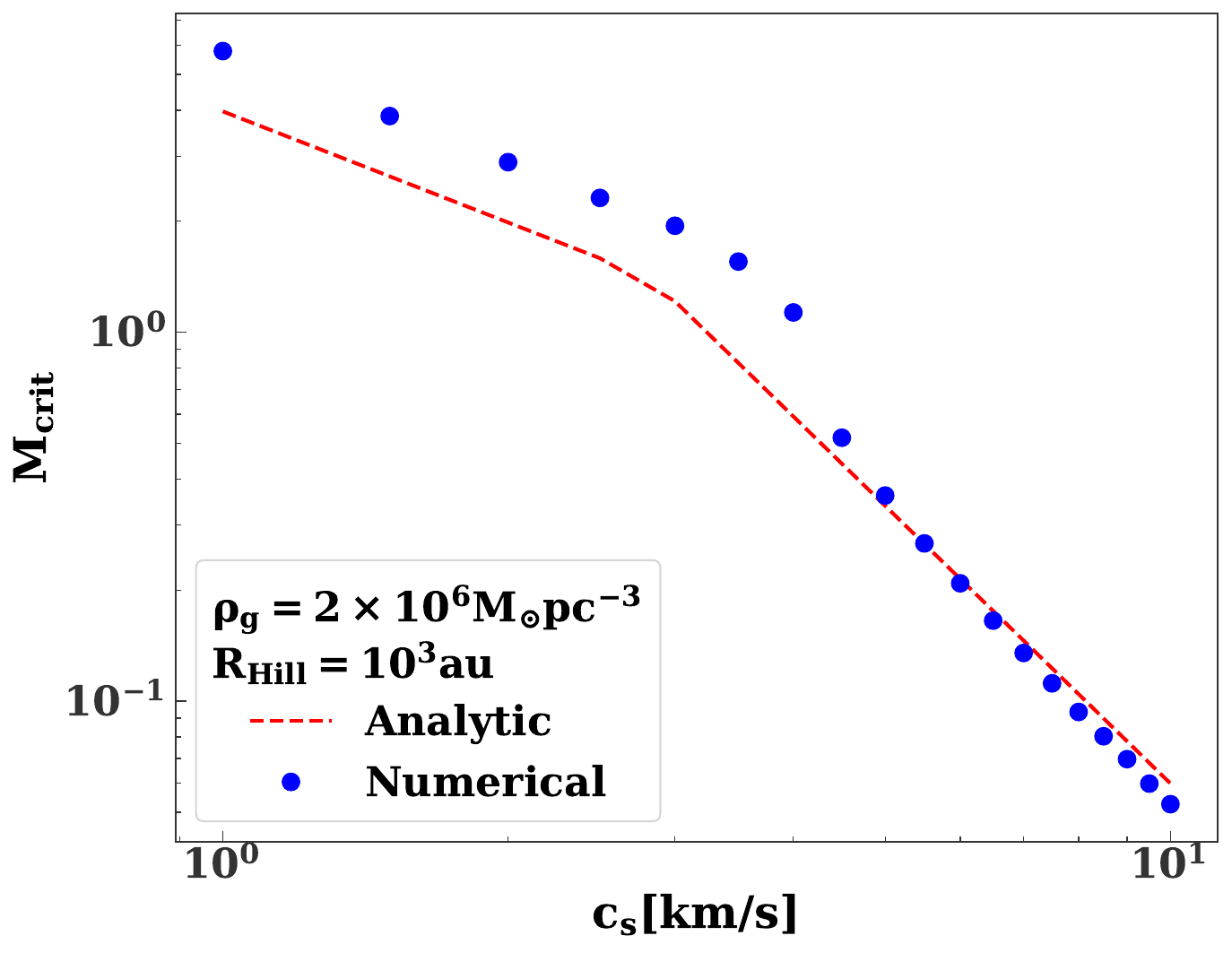}
    \caption{Maximum mach number (at infinity) for which capture can occur as a function of sound speed for different gas densities and Hill radii. The stellar masses are $10 M_{\odot}$. The blue points are from two-body simulations with gas dynamical friction. The red, dashed lines show the analytic estimate for the maximum capture velocity (see equation~\ref{eq:capt} and Table~\ref{table:Critical velocities, vg=0}).}
    \label{fig:captCompare}
\end{figure}

\begin{figure}
    \includegraphics[width=\columnwidth]{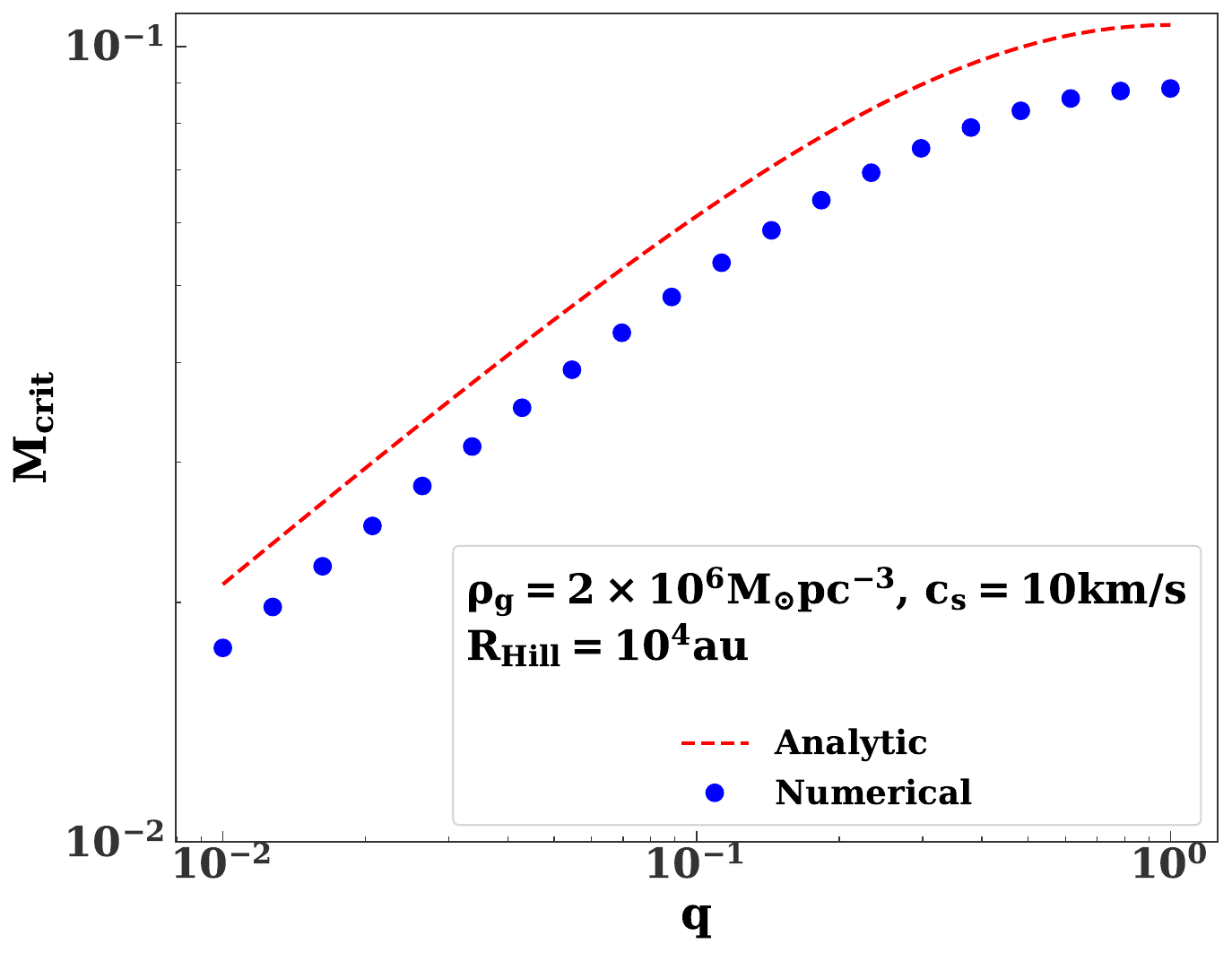}
    \includegraphics[width=\columnwidth]{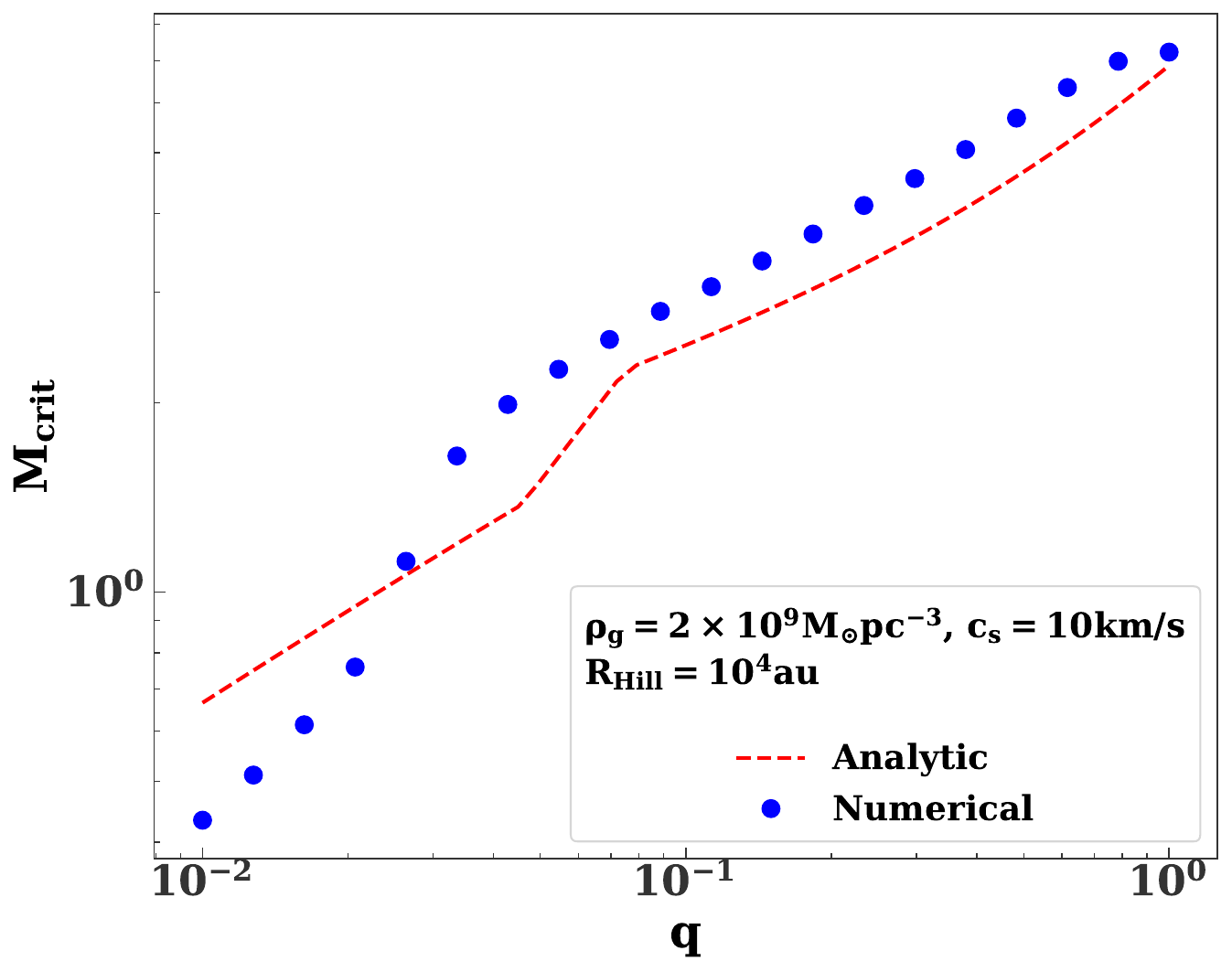}
    \includegraphics[width=\columnwidth]{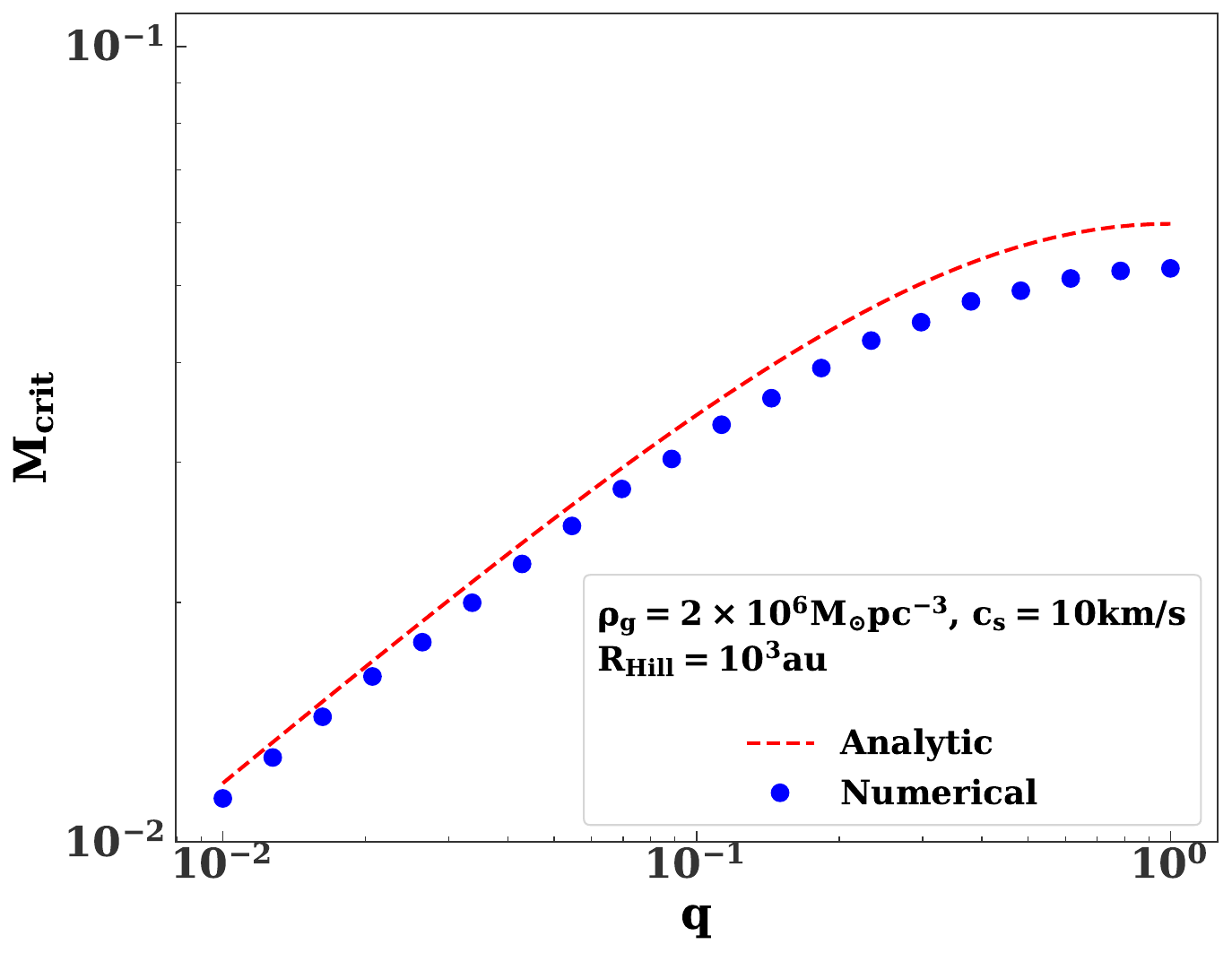}
    \caption{Maximum mach number (at infinity) for which capture can occur as a function of mass ratio for different gas densities and Hill radii. The total mass of the binary is $20 M_{\odot}$. The blue points are from two-body simulations with gas dynamical friction. The red, dashed lines show the analytic estimate for the maximum capture velocity (see equation~\ref{eq:capt} and Table~\ref{table:Critical velocities, vg=0}).}
    \label{fig:captCompare2}
\end{figure}

Figure~\ref{fig:captCompare} shows the maximum mach number for which the two $10 \ M_{\odot}$ stars are captured into a bound binary while crossing the Hill sphere, as a function of the sound speed for a handful of gas densities and Hill radii.\footnote{This is the Mach number at infinity. For each Mach number, the initial velocity corresponds to the velocity at the Hill sphere, accounting for gravitational focusing alone (neglecting the gas).} Figure~\ref{fig:captCompare2} shows the dependence of this capture threshold on the mass ratio of the two stars. For comparison, we also show the maximum mach number from equation~\eqref{eq:capt}. This falls within a factor of $\sim 2$ our numerical results.

\begin{figure*}
    \includegraphics[width=0.32\textwidth]{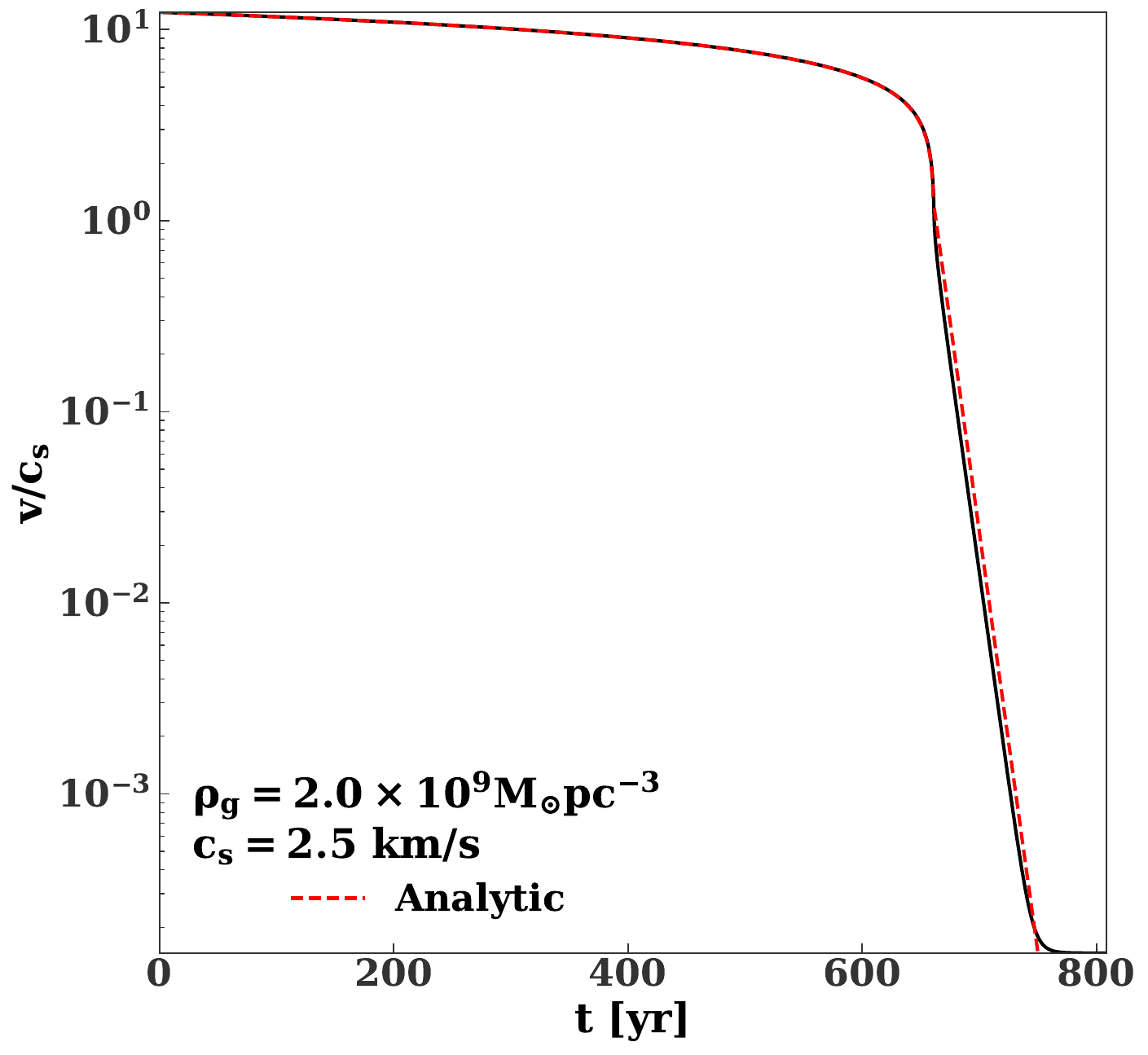}
    \includegraphics[width=0.32\textwidth]{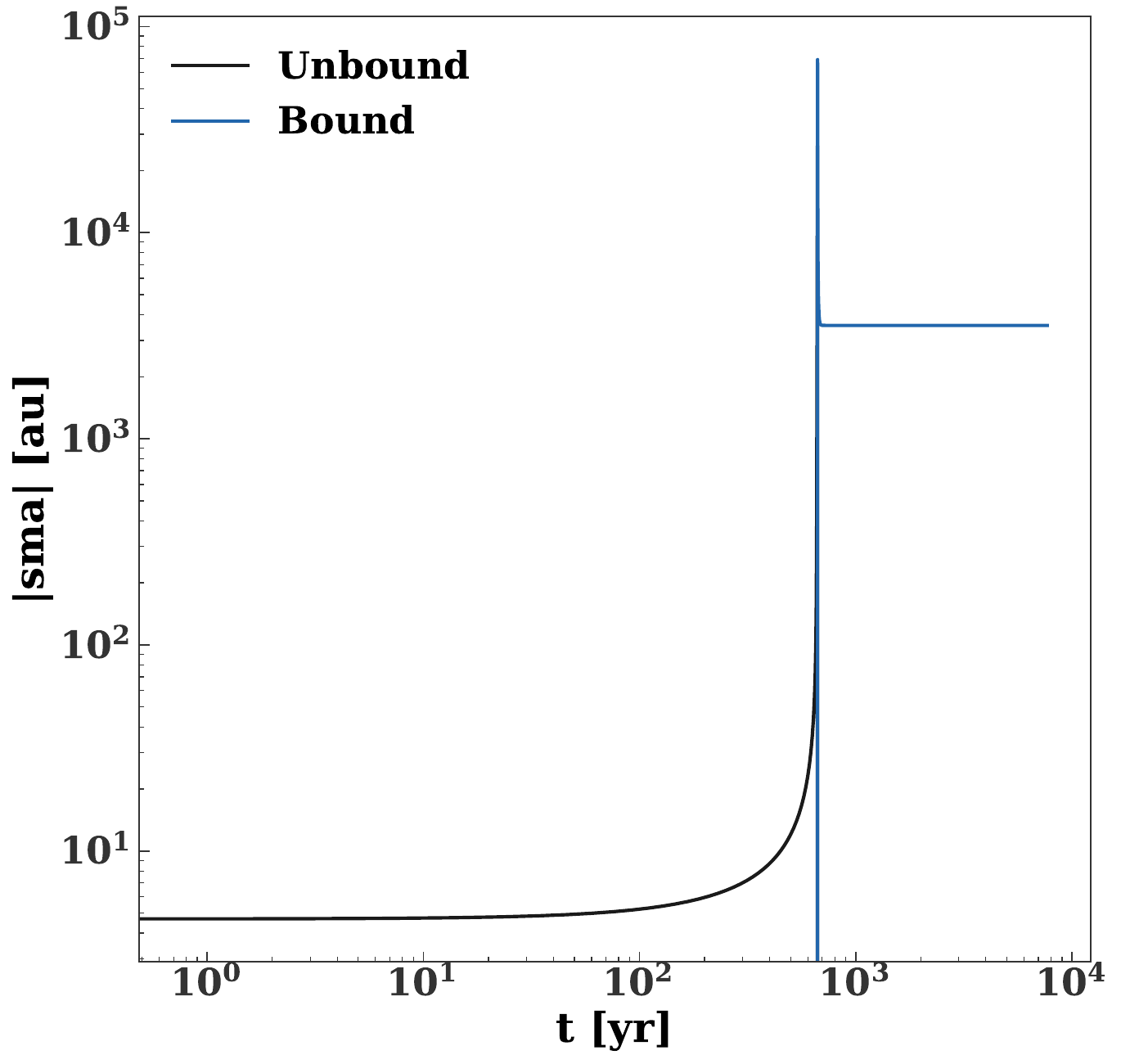}
    \includegraphics[width=0.32\textwidth]{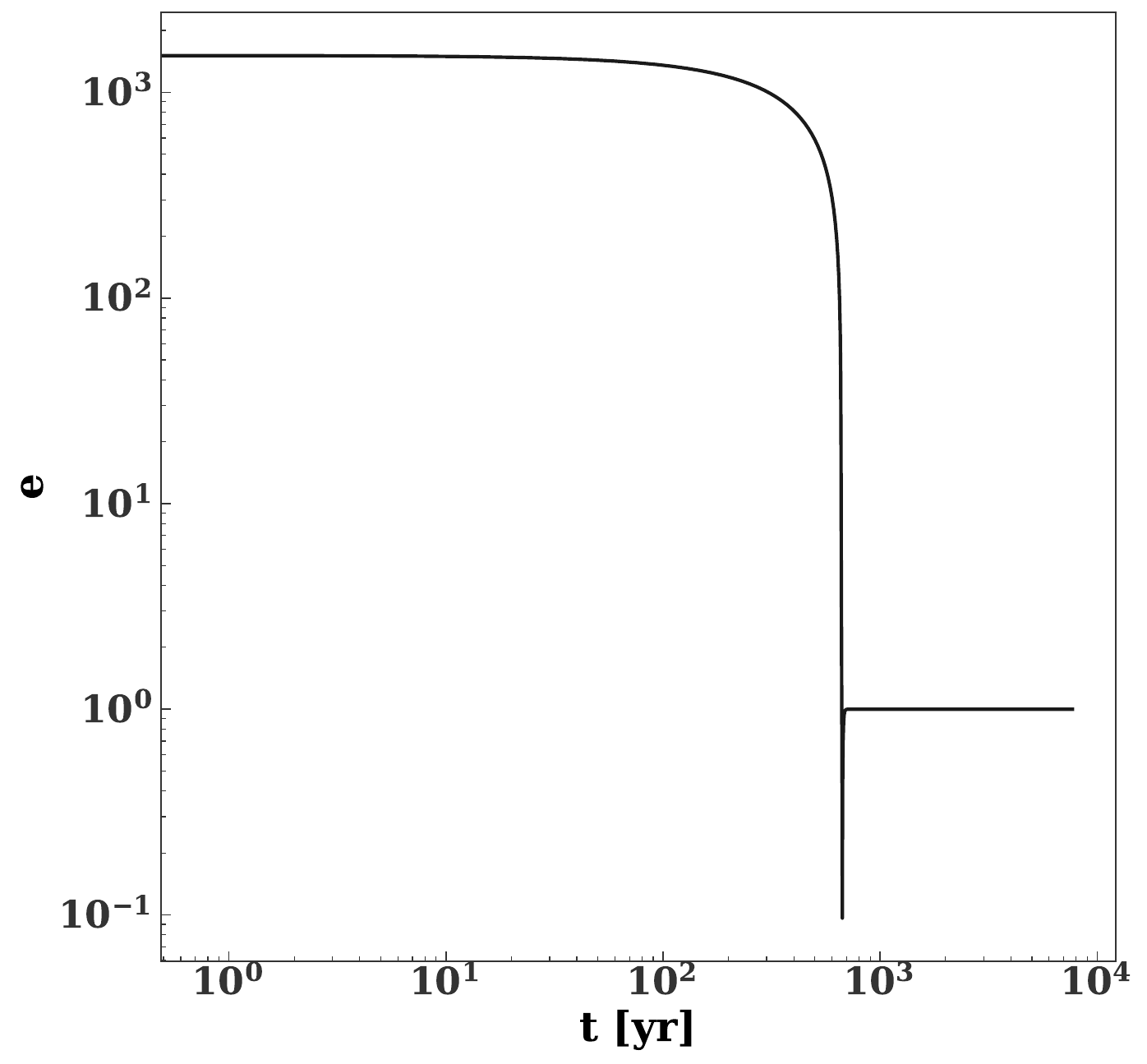}

    \includegraphics[width=0.32\textwidth]{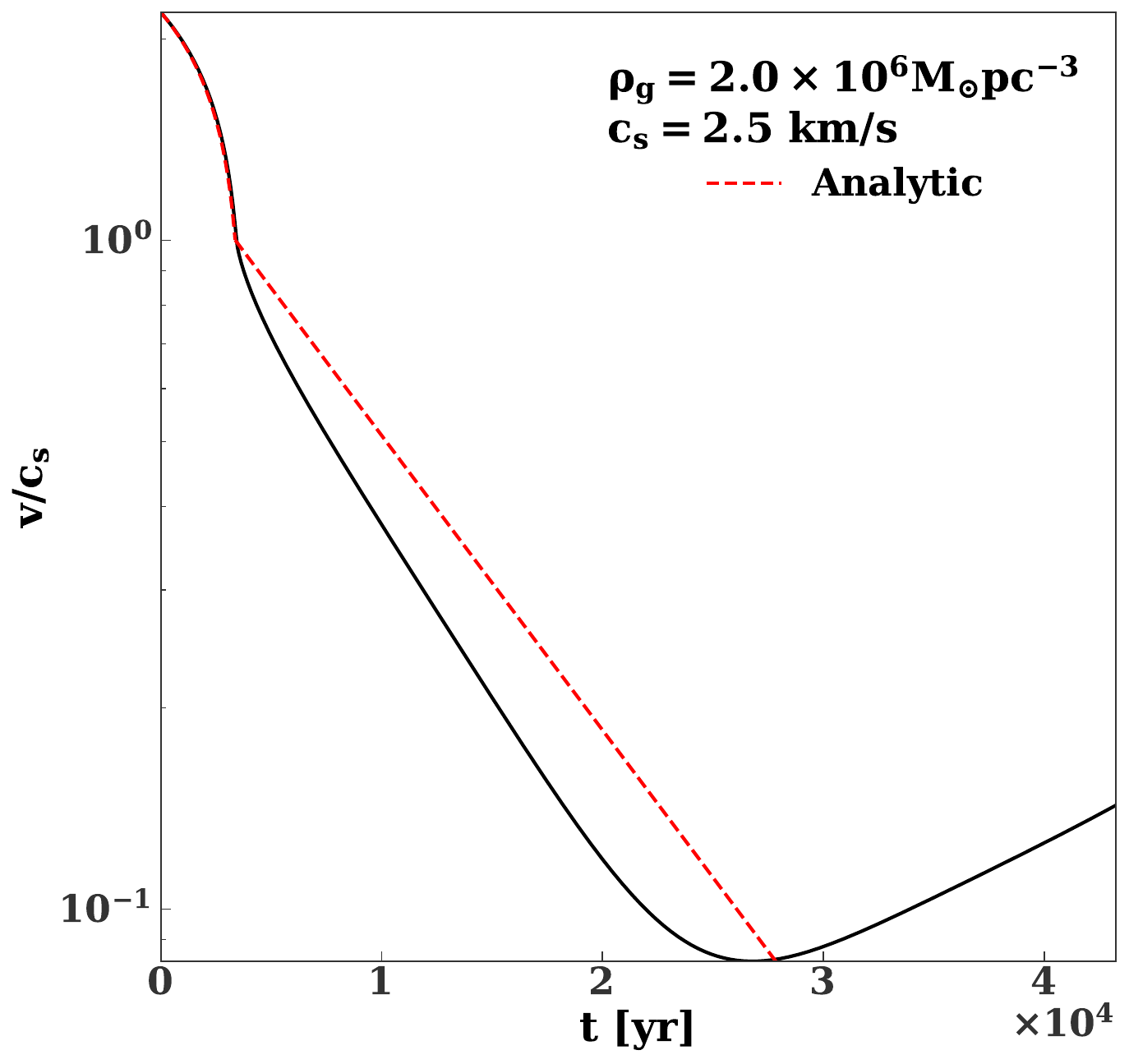}
    \includegraphics[width=0.32\textwidth]{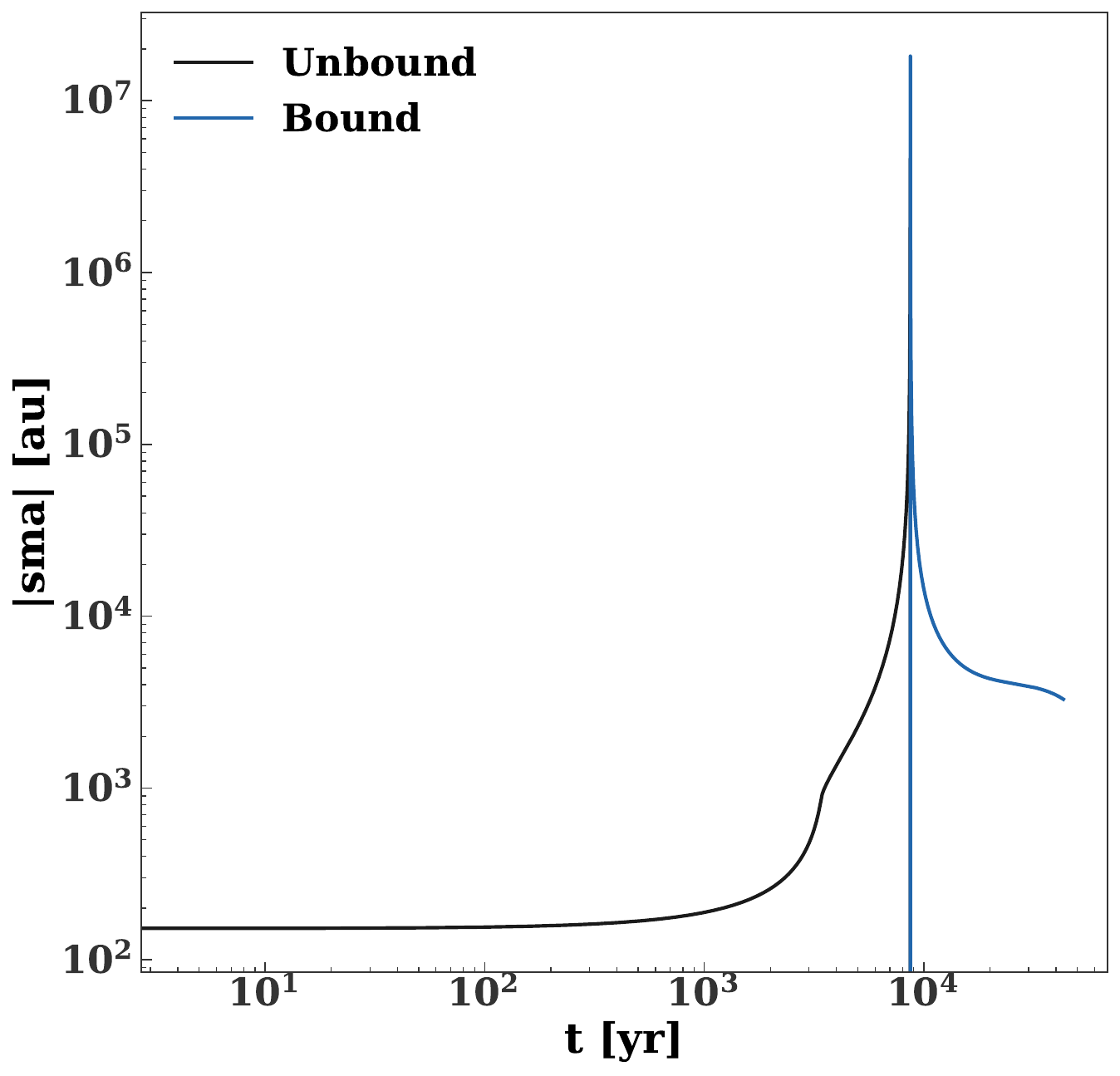}
    \includegraphics[width=0.32\textwidth]{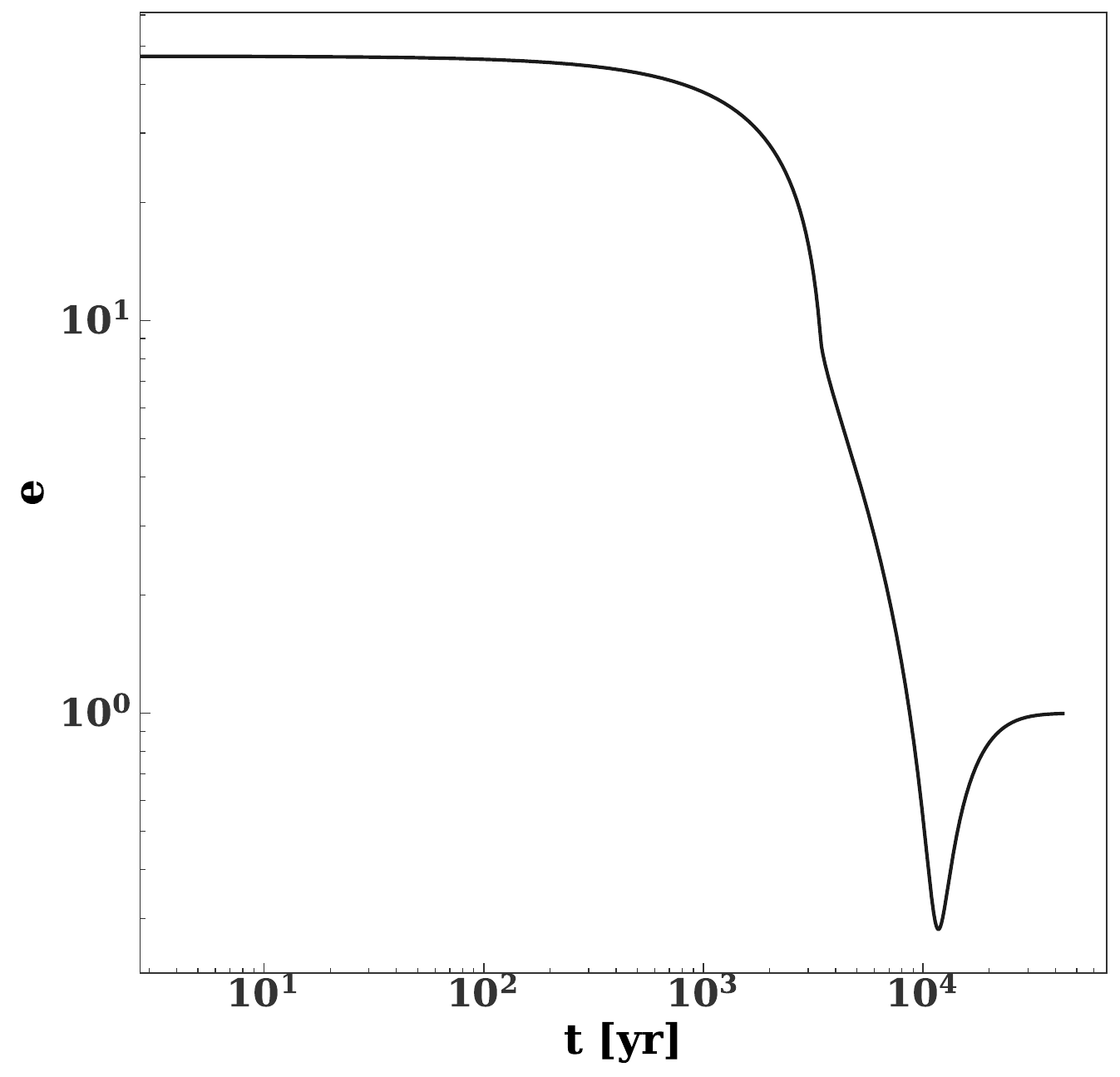}

    \includegraphics[width=0.32\textwidth]{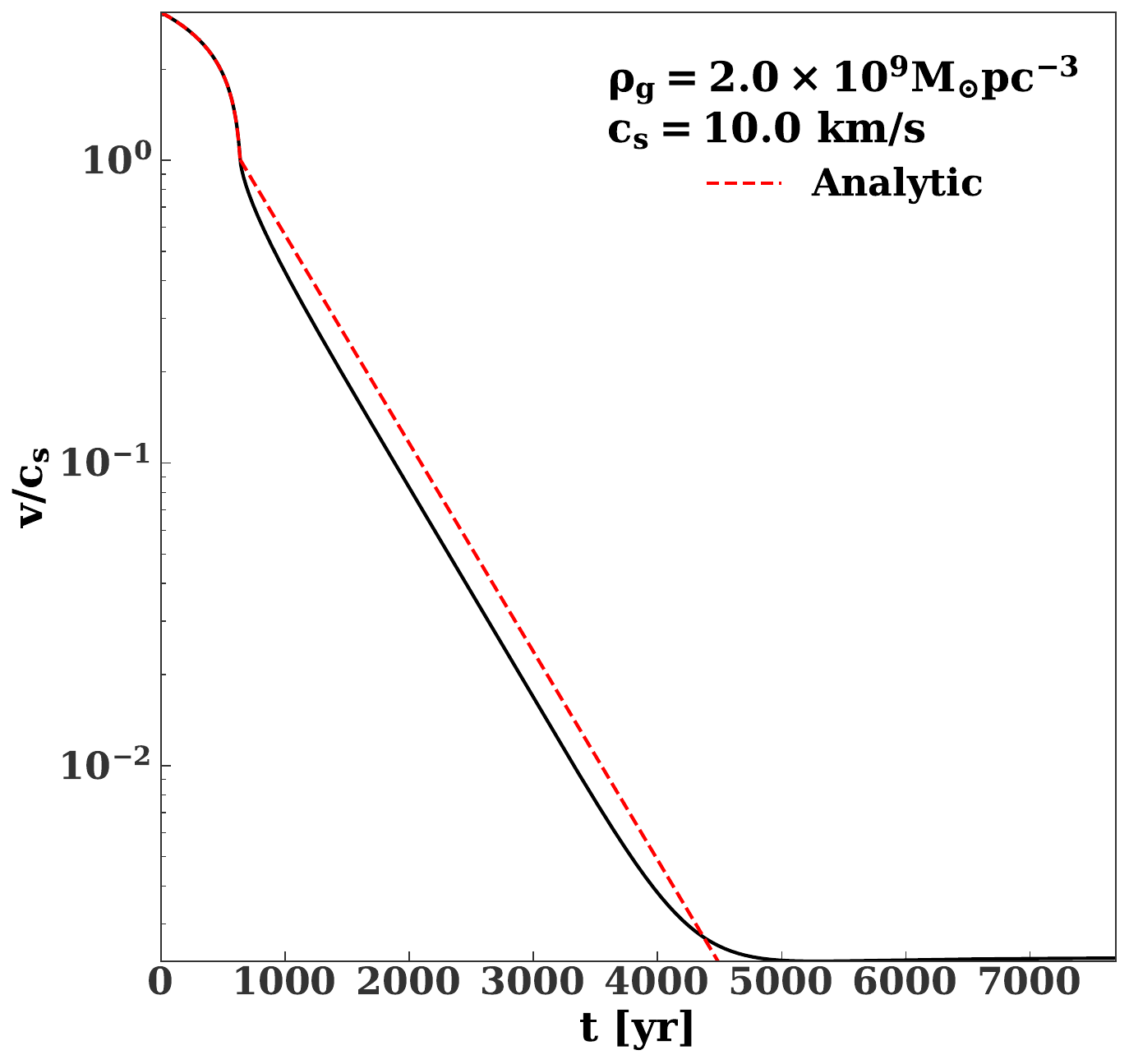}
    \includegraphics[width=0.32\textwidth]{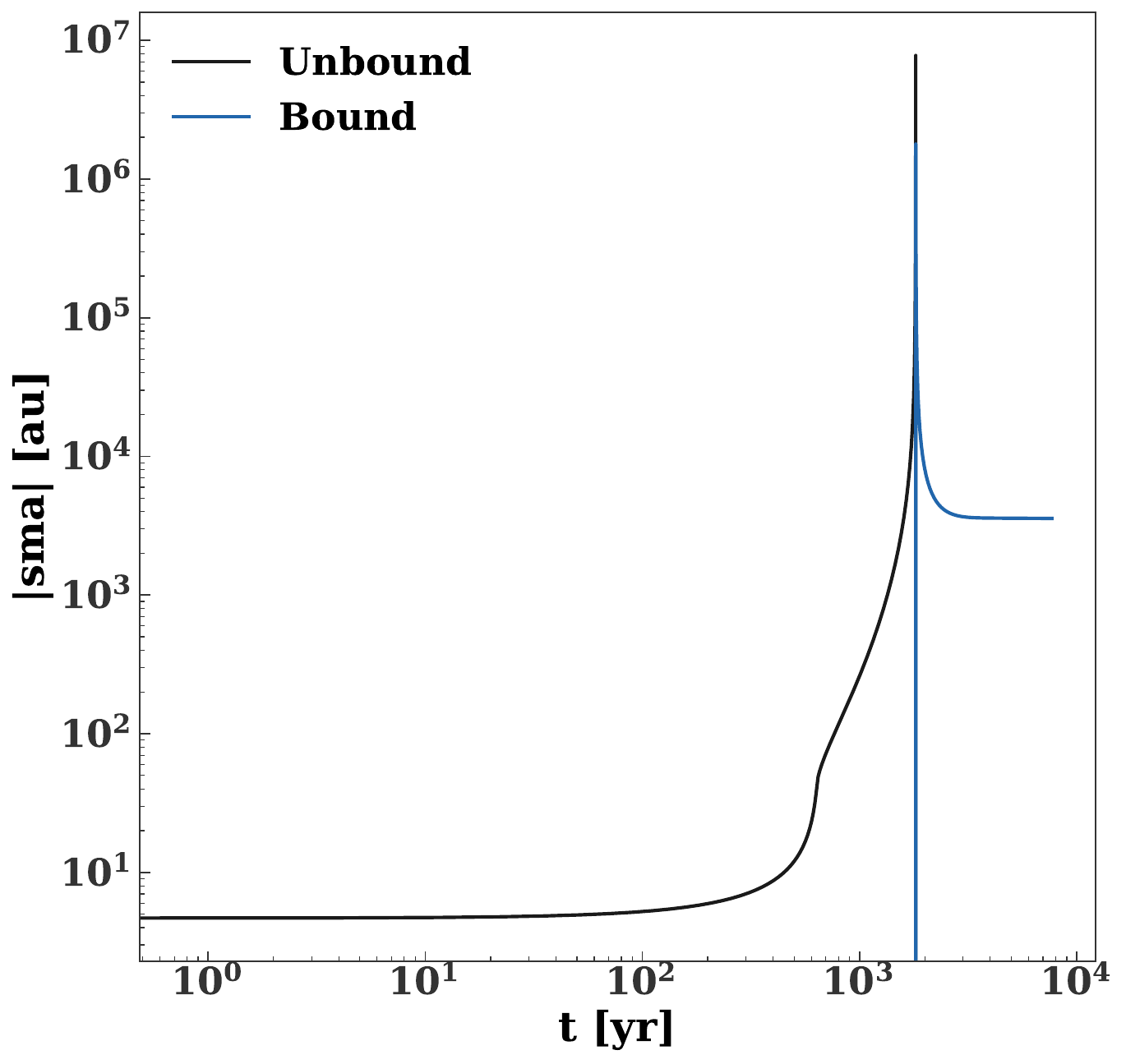}
    \includegraphics[width=0.32\textwidth]{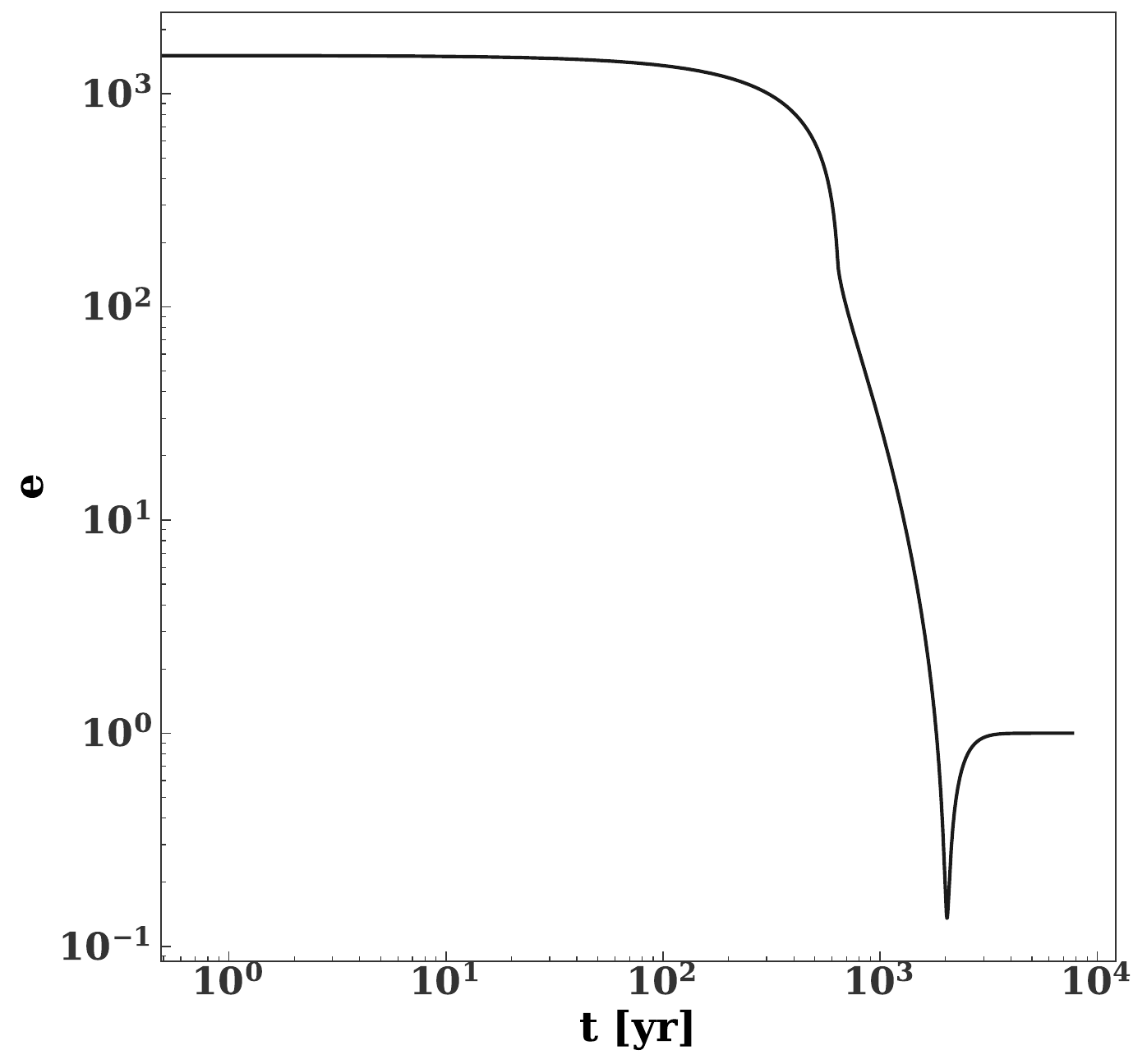}

    \caption{\label{fig:examples} Evolution of the stellar velocities and binary orbital elements from numerical simulations with different gas density and sound speeds. In first column, we also plot the analytic solution of the velocity (from \S~\ref{sec:vel}) as a dashed, red line. In all cases, the stars are both $10 M_{\odot}$.} 
\end{figure*}

\begin{figure}
    \includegraphics[width=\columnwidth]{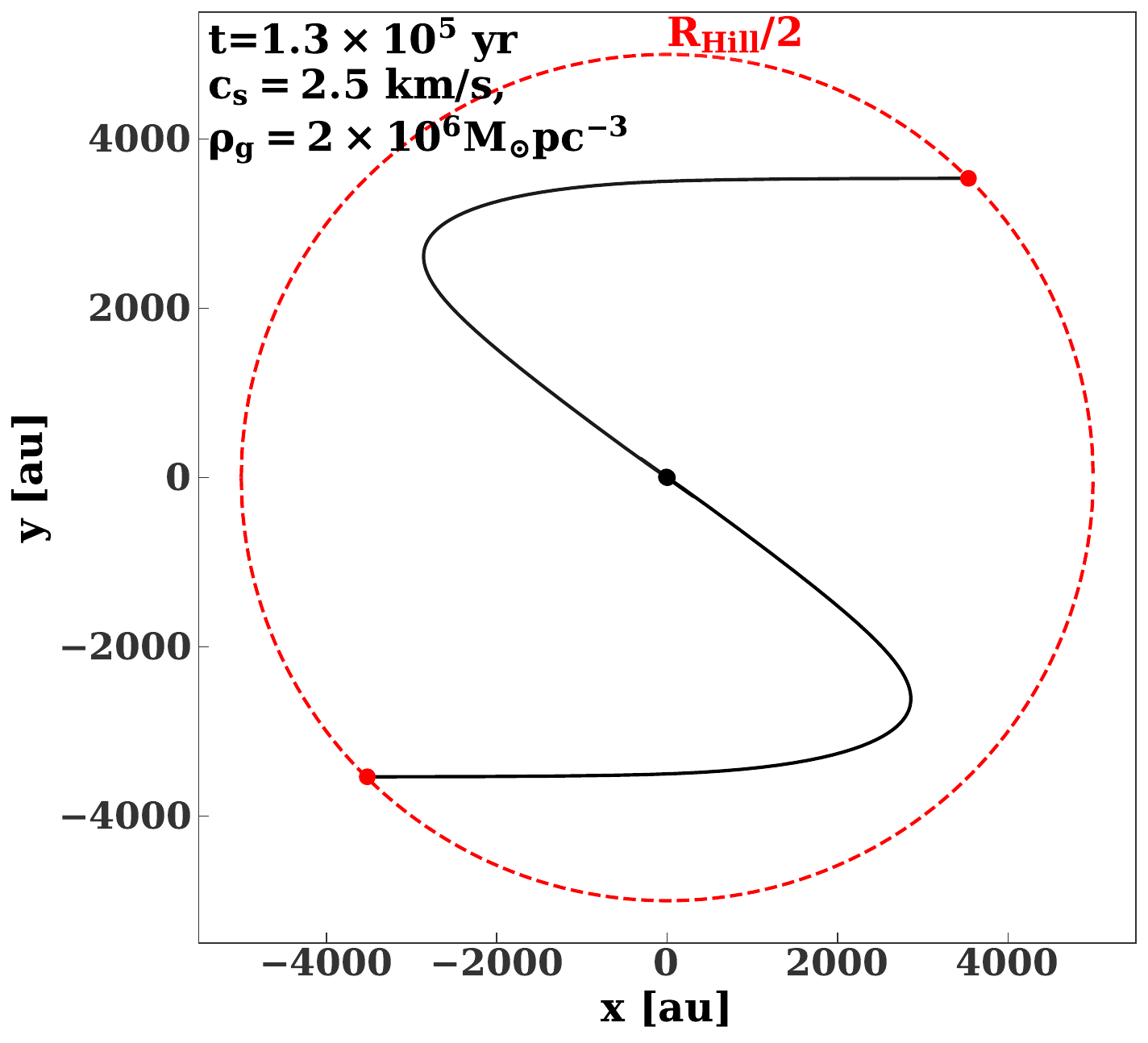}
    \caption{Particle paths for the gas-assisted capture in the second row of Figure~\ref{fig:examples}. The red dots show the initial position of the two $10 M_{\odot}$ objects. The gas is assumed to be at rest with respect to the binary center-of-mass initially. A movie of the capture is available at \url{https://www.youtube.com/watch?v=gh5KC_fjPp4}.}
    \label{fig:merger_example_1}
\end{figure}

Figure~\ref{fig:examples} shows the stellar velocity, binary semimajor axis, and binary eccentricity as a function of time for a handful of gas parameters. We find reasonable agreement between our numerical and analytic solutions for the velocity (see Appendix~\ref{sec:vel}) at early times. At late times, gravitational acceleration (not included in Appendix~\ref{sec:vel}) causes the solutions to diverge. As expected, the binary orbital elements evolve over a dynamical time. Figure~\ref{fig:merger_example_1} shows the trajectories of the stars during the capture in the second row of Figure~\ref{fig:examples}.

 \begin{figure}
     \includegraphics[width=\columnwidth]{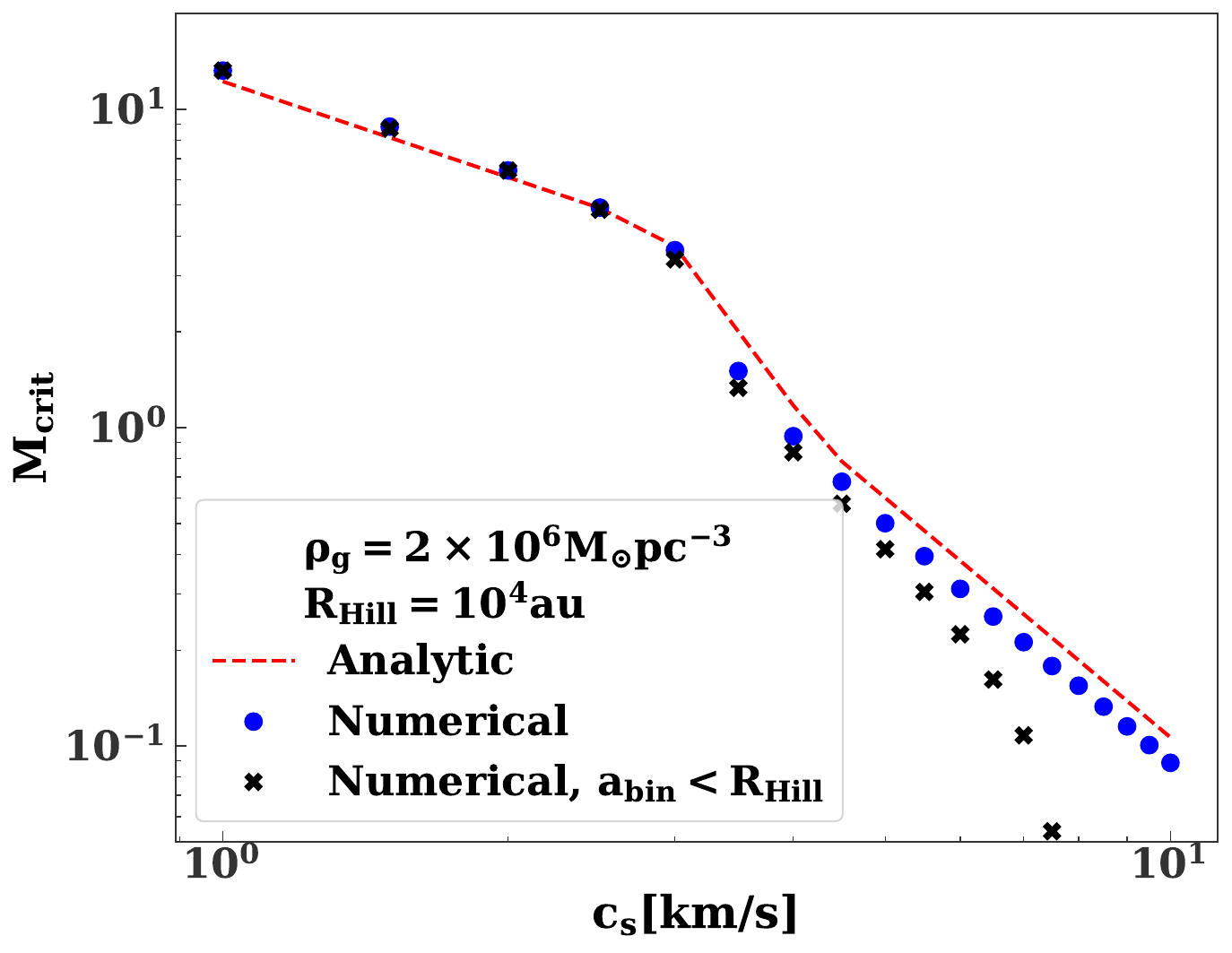}
     \caption{Maximum mach number (at infinity) for which capture can occur as a function of sound speed for the density and Hill radius in the top panel of Figure~\ref{fig:captCompare}. The black crosses show the maximum Mach number for which capture into a stable binary (with semimajor axis less than the Hill radius) can occur. Capture into stable binaries is impossible for sound speeds $\gtrsim 8 \ \rm{km s^{-1}}$.}
     \label{fig:captStab}
 \end{figure}

\subsection{Stability of captured binaries}
\label{sec:focusStability}
If a captured binary forms with a semimajor axis that is greater than the Hill radius, it will be short-lived. Requiring stability can significantly reduce the threshold capture velocity, as shown in Figure~\ref{fig:captStab}.

Captured binaries will necessarily be unstable if the threshold capture velocity is in the gravitationally focused regime.  

 At the threshold velocity, the energy dissipated is precisely the kinetic energy at infinity. If the threshold velocity is in the focused regime this will be small compared to the potential energy at the Hill sphere. Furthermore, if the encounter is focused, the relative velocity of the object in the relevant environment (e.g. star/compact-object or Kuiper-belt object; in cluster/AGN or solar-system environment, respectively) inside the Hill sphere (and hence the energy dissipated) is a weak function of the velocity at infinity. Thus, the energy dissipated will always be small compared to the potential energy at the Hill sphere in this regime and no stable binaries can form.
 
\subsection{Effects of a headwind}\label{subsec:headwind}

So far we have neglected the binary's center-of-mass motion through the gas. However, there might be a non-negligible headwind that could affect the capture and subsequent evolution. We perform additional few-body simulations with a headwind, and discuss the results here.

\begin{figure*}
    \includegraphics[width=0.99\columnwidth]{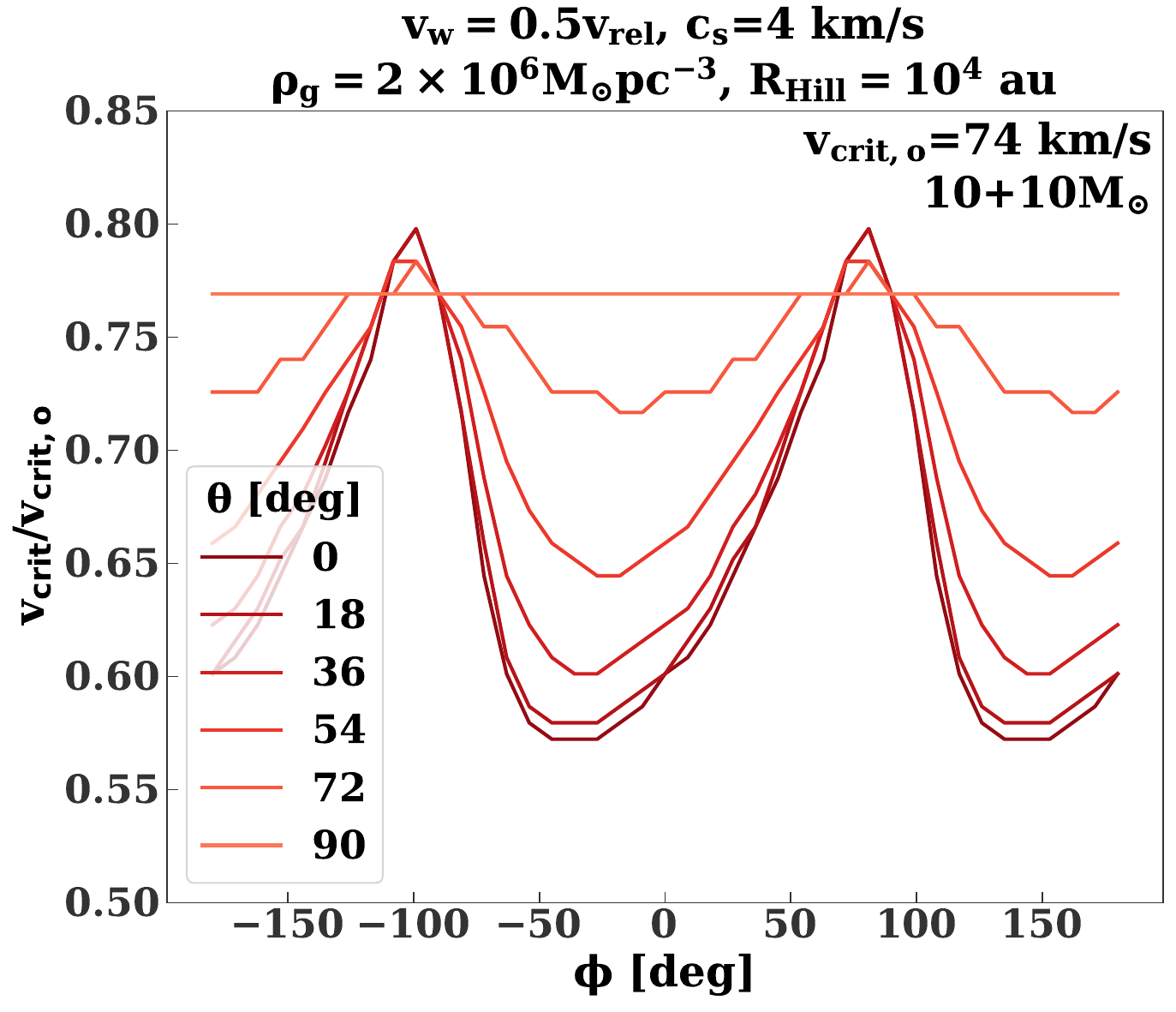}
    \includegraphics[width=0.99\columnwidth]{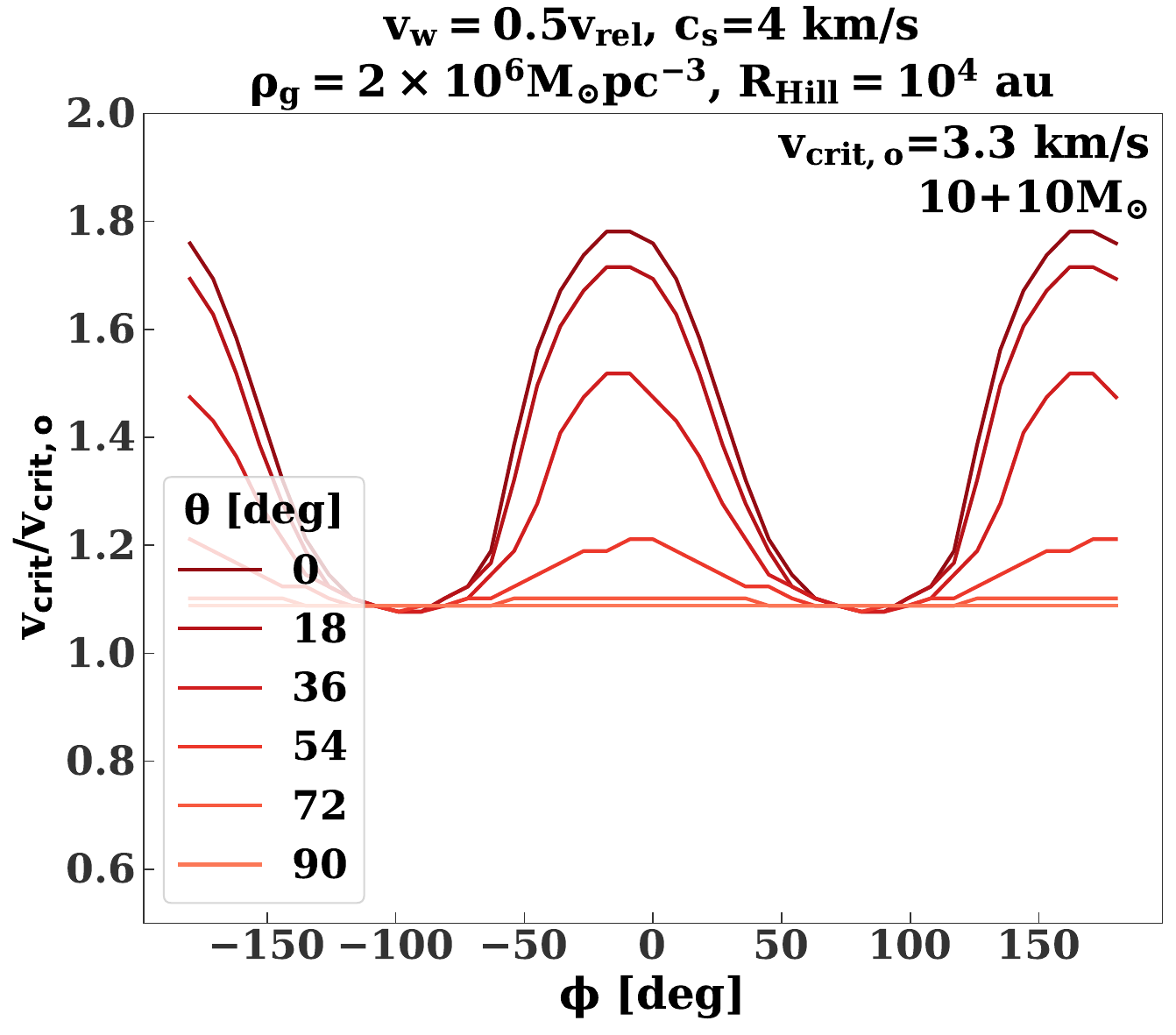}
    \includegraphics[width=0.99\columnwidth]{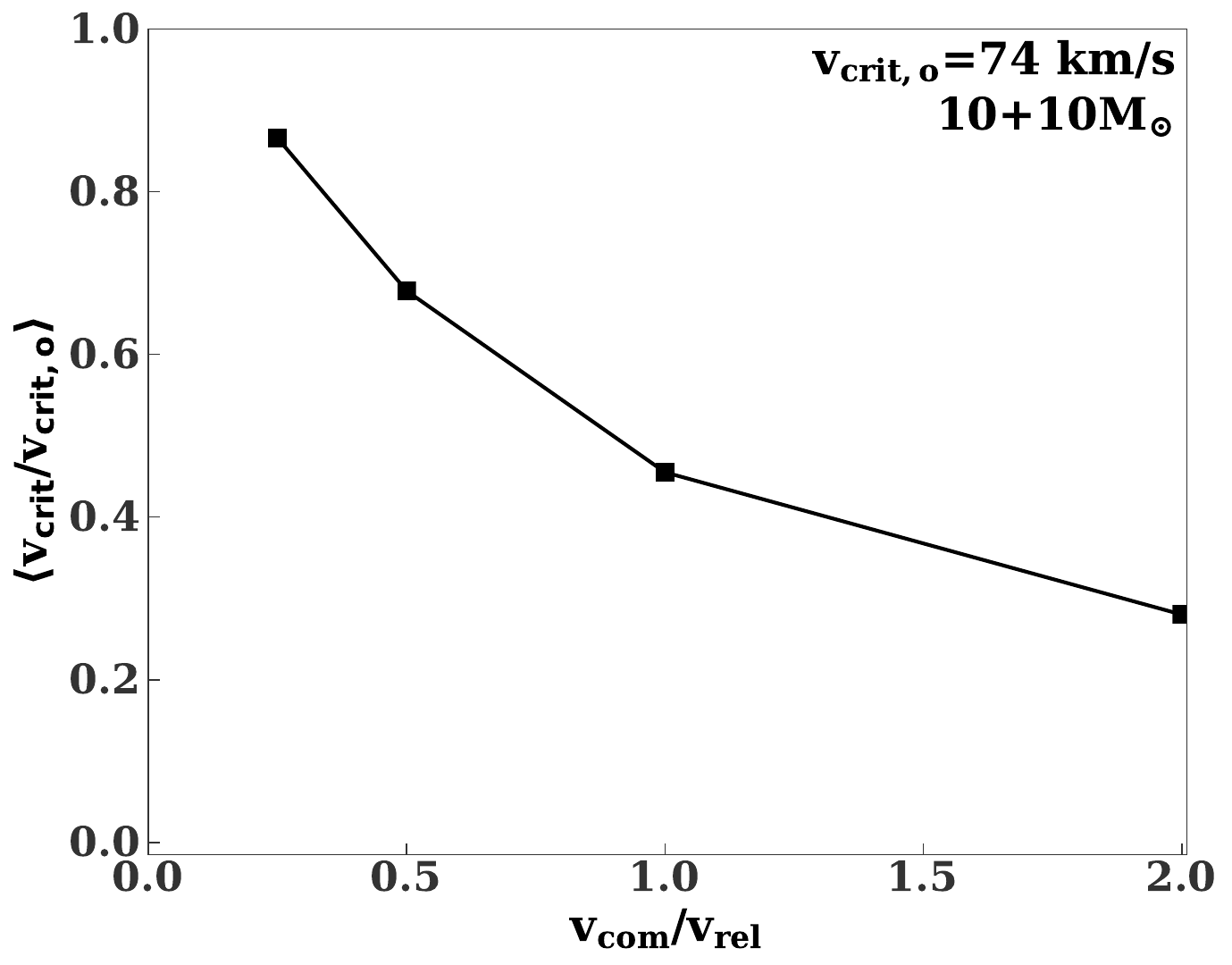}
    \includegraphics[width=0.99\columnwidth]{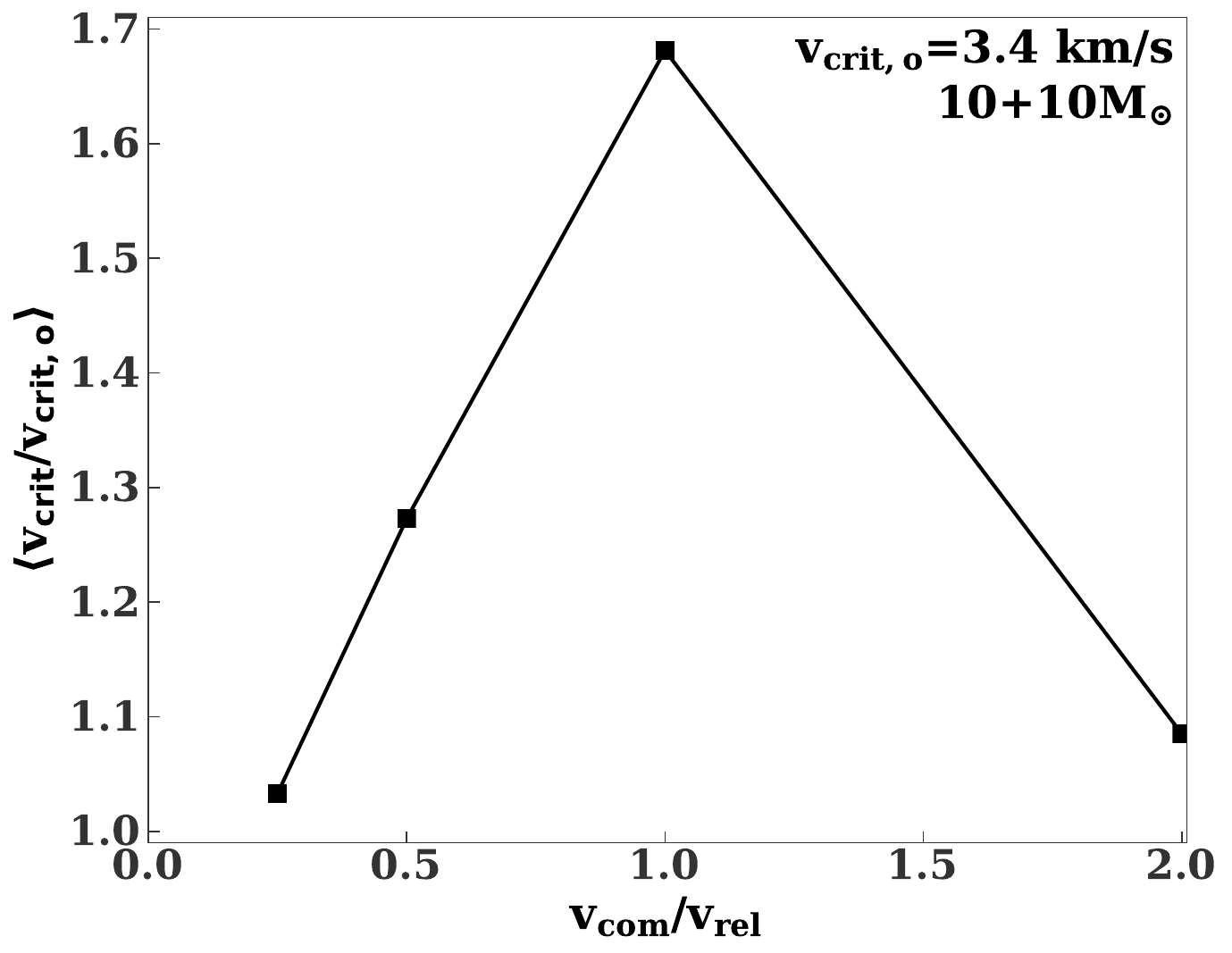}
     \caption{Two examples of the effect of binary motion on the threshold capture velocity. The left (right) panels correspond to the supersonic (subsonic), unfocused regime. Top panels shows the change in the threshold capture velocity as a function of the direction of motion. Here, the bodies' center-of-mass velocity (with polar angles $\theta$ and $\phi$) is half the relative velocity.
     The bottom panels show the angle-averaged ratio between the thresholds with and without the center-of-mass motion as a function of the center-of-mass velocity (normalized to the relative velocity). } 
    \label{fig:headWindSup1}
\end{figure*}

Figure~\ref{fig:headWindSup1} shows the effect of this motion on the threshold capture velocity. In the unfocused, supersonic regime it reduces the threshold. For large center-of-mass velocities, the reduction is significant. (It is up to a factor of $\sim$4 if the center-of-mass velocity is twice the relative velocity, though this precise reduction depends on the orientation of the headwind). However, for two equal mass objects with isotropic, Maxwellian velocities the center-of-mass velocity will be approximately half the relative velocity on average. In this case, the threshold is reduced by a factor of 0.68 on average. In the unfocused, subsonic regime the center-of-mass motion increases the threshold velocity and aids capture. 

Typically, the binaries center-of-mass motion will affect the threshold for capture by less than a factor of 2. Thus, we neglect this effect in our capture rate estimates.

\section{Capture rates}

For a given environment, the binary formation rate through gas-assisted capture could be written by  

\begin{align}
& \Gamma(m_1,m_2) \approx \int_0^{v_{\rm crit}}n_\star(m_2|m_1) \mathcal A v p(v)dv, \nonumber\\
&\mathcal A =
R_{\rm Hill} z(1+\Theta^2)
\nonumber\\
& z = \min\{R_{\rm Hill}, h_{\rm eff}\}
\label{eq:captureRate}
\end{align}
where
$m_1$ is the mass of the capturer, $m_2$ is the captured mass, $n_\star(m_2|m_1)$ is the density of candidates for captured masses in the vicinity of $m_1$, $h_{\rm eff}$ is the effective scale height of the disk (if the environment has a disk-like configuration; e.g. an AGN disk, a gaseous disk in a cluster or a protoplanetary disk), $p(v)$ the velocity distribution and $\Theta=(v_{\rm{esc}}/v)^2$ is a correction for gravitational focusing. This correction is not valid in systems dominated by a massive central object like AGNs and protoplanetary discs, since it is derived assuming unperturbed two-body trajectories and neglects shearing motion. 
The critical velocity for capture, $v_{\rm crit}$, is calculated according to the regime (focused/unfocused), as specified in Table \ref{table:Critical velocities, vg=0}.
We assume this is a Maxwellian distribution, such that $p(v)\propto v^2 e^{-v^2/2\sigma^2}$, where $\sigma$ is the velocity dispersion. 
Thus, equation~\eqref{eq:captureRate} simplifies to 

\begin{align}\label{eq:capturare_unfocused}
&\Gamma(m_1,m_2)=
n_\star R_{\rm Hill} z \sigma  \sqrt{\frac{2}{\pi}} \left[f_1 +f_2\right]\nonumber\\
&f_1=\left(v_{\rm esc}^2/\sigma^2\right) \left(1-e^{-v_{\rm crit}^2/(2 \sigma^2)}\right)\nonumber \\
&f_2=2-e^{-v_{\rm crit}^2/(2\sigma^2)} \left(2+\left(\frac{v_{\rm crit}}{\sigma}\right)^2 \right)
\end{align}
Note that this expression differs from the one derived for example in \cite{Tagawa2020} by the factor in the brackets (divided by $2\sqrt{2\pi}$). This term becomes significant if $v_{\rm crit}<\sigma$, where capture is dominated by the tail if the velocity distribution.

For small capture velocities, the capture for the unfocused case rate could be approximated by 

\begin{align}
\Gamma(m_1,m_2)\approx 
\sqrt{\frac{2}{\pi}}\frac{n_\star R_{\rm Hill}^2 v_{\rm crit}^4}{2\sigma^3}
\end{align}
Below we derive the gas-assisted capture rate for different gaseous environments and set constraints on the available parameter space that enables such a capture. 
Throughout this paper, we use eq.~\eqref{eq:captureRate} to calculate the capture rate, dropping the focusing correction in shear-dominated environments (AGN and protoplanetary discs).

\section{Capture rates in different gas-rich environments}
\label{sec:discussion}
In this section, we study the conditions for binary formation in different environments, and summarize our results in Table \ref{table:capture rates}. We consider  (1) Star-forming environments where binaries are formed from newly born stars or even pre-main-sequence stars or protostars, where gas-assisted capture may serve as an important channel for the fundamental formation of stellar binaries; (2) AGN disks around supermassive black holes, where gas-assisted captures could form stellar and compact objects binaries, and may contribute to the formation of stellar binaries which could later inspiral and eventually give rise to merger products and explosive transients and GW sources from compact-object mergers; (3) Gas-enriched massive clusters, where the existence of multiple generation of stars suggest several epochs of gas rich environments in which earlier generations of stars and compact objects could be  embedded, and form stellar and compact-object binaries, similar to the case of AGN disks. (4) Gaseous protoplanetary disks where embedded planetesimals can form binaries through gas-assisted capture, and in particular, the early stages of planet formation in the Solar systems could give rise to the production of Kuiper-belt objects (KBO) and asteroid binaries.    

\begin{table*}
  \centering
\begin{tabular}{
|p{2cm}|p{1.5cm}|p{1.5cm}|p{1.5cm}|p{1.5cm}|p{1.5cm} |p{1.5cm}|p{1.5cm}
}
 \hline
 \hline
& $n_g [ \rm{cm^{-3}}]$
& $m_{\rm bin}[M_\odot]$&$R_{\rm Hill}[\rm{AU}]$ & $c_s[\rm{km/sec}]$ & $n_{\rm b}[\rm{pc^{-3}}]$  & $\Gamma[\rm{Myr^{-1}}]$   \\
\hline
 \hline
 SF   &   $10^4$  & $2$ &  
 $5 \times 10^4$
 & $0.2$ & 
 $308$
 &
 $12.6$
 \\
 \hline
later SF 
-- GCs & $4\times 10^6$ & $20$ & $1.2\times10^4$ & $0.6$ &$10^3$ &  
$9.2$
\\
 \hline
 AGN$^*$ & $4\times 10^7 $ & $20$& $2.4\times 10^3$ & $0.4$ & $9.5\times 10^4$ & 1 \\
 \hline
 PPD & $4\times 10^{11}$ & $4\times 10^{-12}$ & $6\times 10^{-3}$ & $0.15$ & $ 3\times 10^{15}$ & 
$\ll (\rm{gas \ lifetime})^{-1}$
 \\
 \hline \hline
\end{tabular}
\caption{ Typical capture rates per object, for equal mass binaries, in four different environments: SF regions, later generation formation in GCs, AGN disks and protoplanetary disks. The columns refer correspondingly to the gas number density $n_g$, binary mass $m_{\rm bin}$ for which the capture rates are presented, Hill radius $R_{\rm Hill}$, sound speed $c_s$, background density of the captured objects $n_b$ and finally the expected capture rate $\Gamma$, as calculated according to our model, taking into consideration the relevant regime of focusing. Note that the rate for AGN disks is averaged over a non-flat stellar density profile. \label{table:capture rates}}
\end{table*}

\subsection{Star forming environments}\label{subsub:SF}

Star formation takes place in 
cold gas-rich clumps embedded in molecular clouds. These clumps could constitute as a fertile ground for gas-assisted binary formation. 
The typical gas temperature in these regions is $\sim 10 \ K$ \citep{Shu1987,Williams2000}, which corresponds to a sound speed of $0.2 \ \rm{km/sec}$. The typical mass of clumps is $10^3-10^4 \ M_\odot$ and their radii are $2-5 \ \rm{pc}$ \citep{Shu1987}.

The typical gas density in the clumps should exceed a threshold value to enable star formation, which is typically $n_{\rm th} \sim 10^4 \ \rm{cm}^{-3}$ (\citealp{BerginTafalla2007} and references therein). 
Assuming the mass in stars is comparable to the mass in gas, the gas and stellar densities could be approximated by
$\rho_{\rm stars}=\rho_{\rm gas}=n_g m_H\approx 225 \ M_\odot \ \rm{pc}^{-3}$.

Unless stated otherwise, the radius of the clump is $2 \ \rm{pc}$, the clump mass is $10^3\ M_\odot$, and the gas density is $n_g=10^4 \ \rm{cm^{-3}}$. Other parameters are derived from these choices.
For the stars, we assume Kroupa mass function \citep{Kroupa2001}. 
Using these parameters, we calculate the rate of gas-assisted binary captures using equation~ \eqref{eq:captureRate}.

\begin{figure*}
    \includegraphics[width=\columnwidth]
    {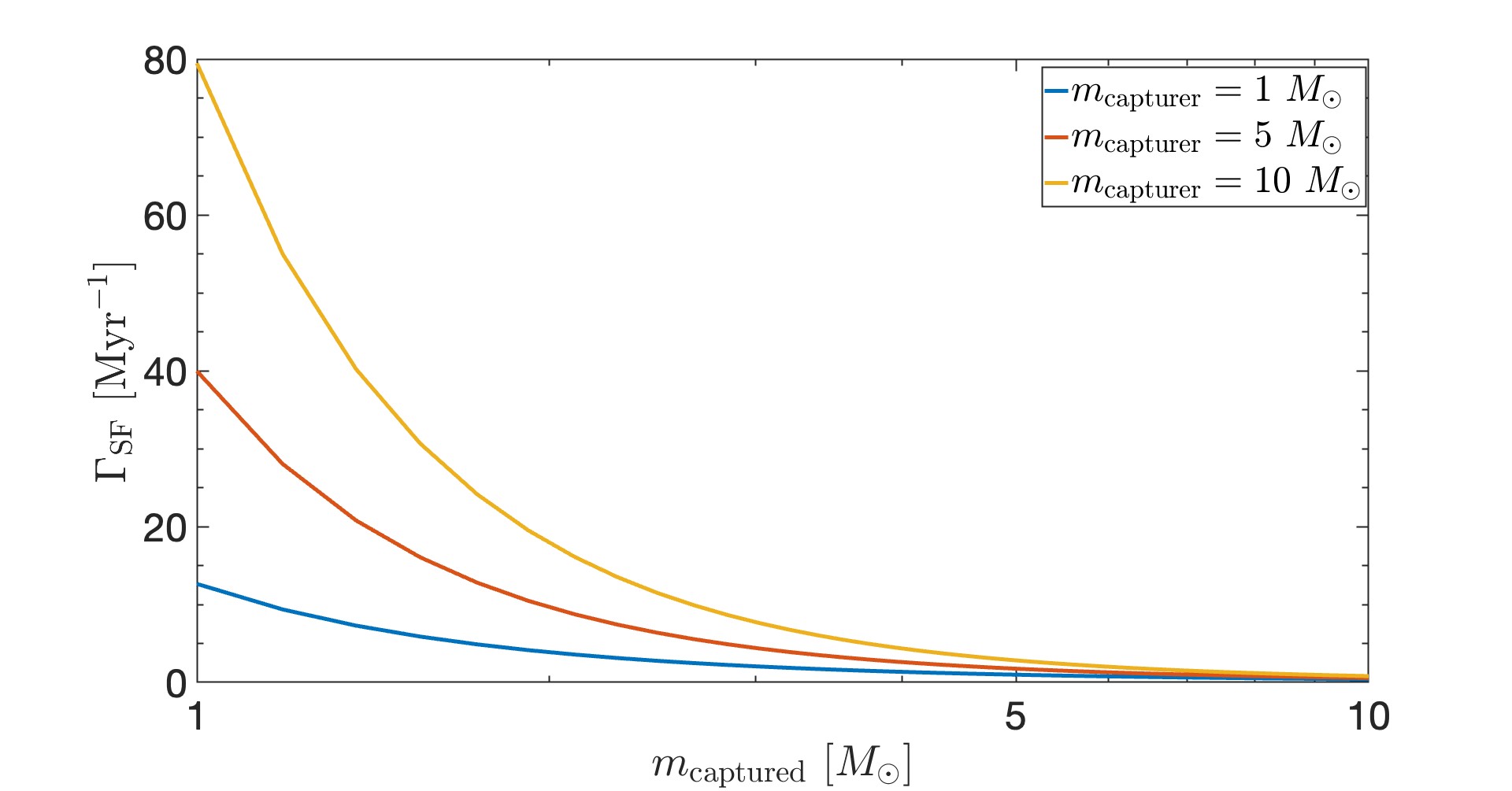}
    \includegraphics[width=\columnwidth]
    {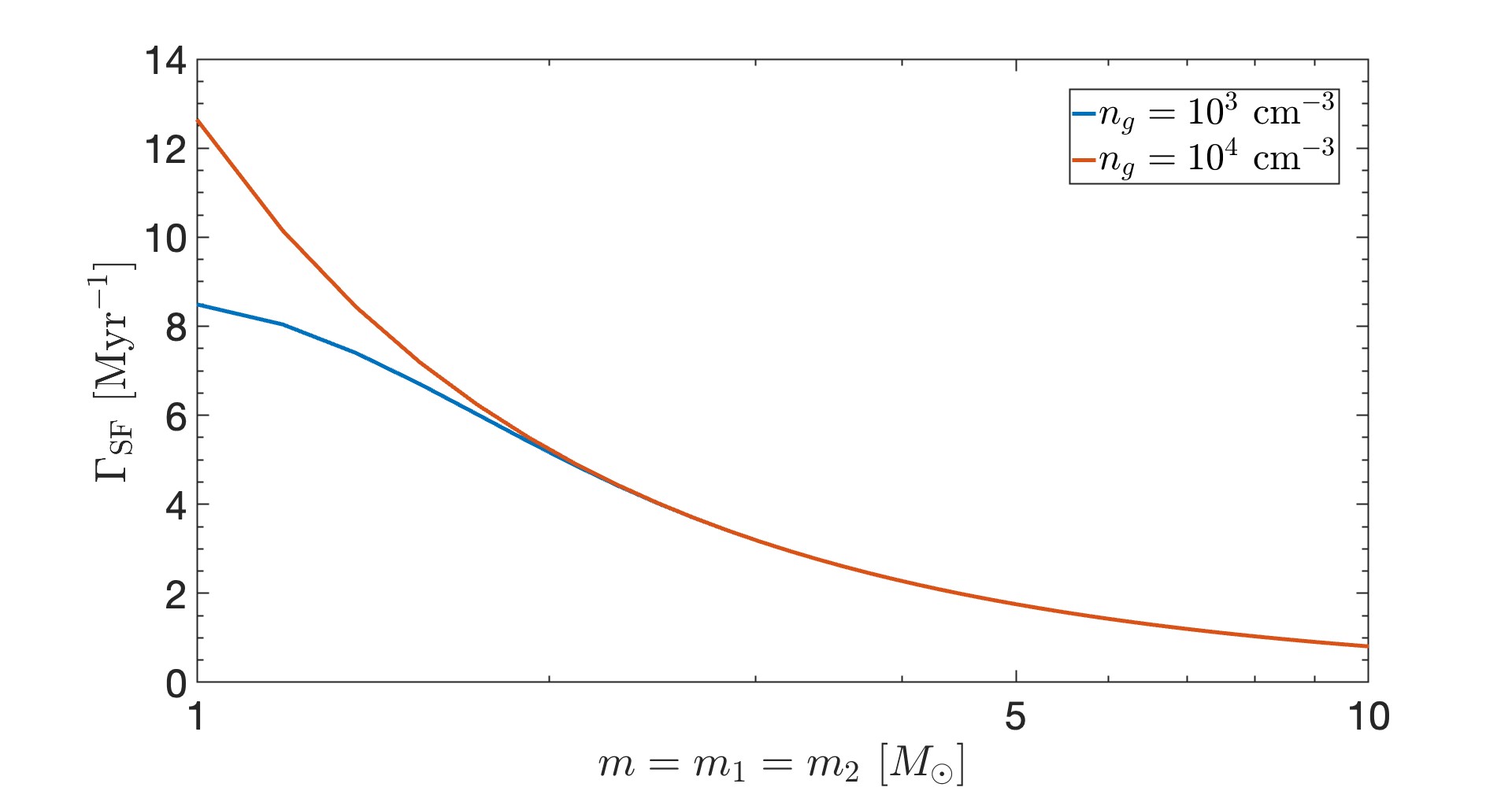}
    \caption{Left: The capture rate per object for different masses, in a SF environment, for our fiducial model specified above. Right: The capture rates for equal masses, given different gas number densities.  \label{fig:SF}}
\end{figure*}

In Figure~\ref{fig:SF}, we present the capture rate per object in a SF region. The gas-assisted binary formation is robust and every object in this environment is likely to capture at least another object during the gas lifetime, and even more. As expected, the capture rate increases with the gas density. It should be noted that the capture rate presented corresponds to $m_1$ capturing $m_2$ -- i.e. when calculating the rate (eq.~\ref{eq:captureRate}), the background number density changes with the captured species according to the background mass function. To calculate the total number of binaries with masses $m_1$ and $m_2$, one should sum up the contributions from $m_1$ capturing $m_2$ and vice versa.

Further evolution of these formed binaries is left out for future studies (in prep.), and could leave unique signatures on binaries distributions. It should be noted that past studies already discussed orbital decay of binaries due to gaseous background in similar context (e.g. \citealp{Stahler2010,Korntreff2012})
\subsection{AGN disks}\label{subsub:AGN}
The evolution of binaries in AGN disks were studied extensively (e.g. \citealp{McKernan2012,Stone2017,Tagawa2020} and references therein), especially as progenitors for GWs.
Gas-assisted inspirals were modeled in different ways, either through a planetary migration modeling \citep{McKernan2012,Stone2017}, gas dynamical friction (e.g. \citealp{Bartos2017}), or Bondi-Hoyle accretion \citep{AntoniMacLeodRamirezRuiz2019}, and was also explored explicitly through hydrodynamical simulations (e.g. \citealp{AntoniMacLeodRamirezRuiz2019,LiLai2022} and references therein)
, though the actual migration modeling is still debated.
It was suggested by \cite{Tagawa2020}, that the vast majority of merging binaries in AGN disks originate in gas-assisted binary formation. Hence, the conditions for binary formation effectively dictate the expected merger rates in such environments. There is a wide range of possible AGN configurations and masses of the central massive black holes (MBHs). We will consider a more specific case, but the same approach could be generalized to other AGN conditions. 

\begin{figure*}
    \includegraphics[width=\textwidth]{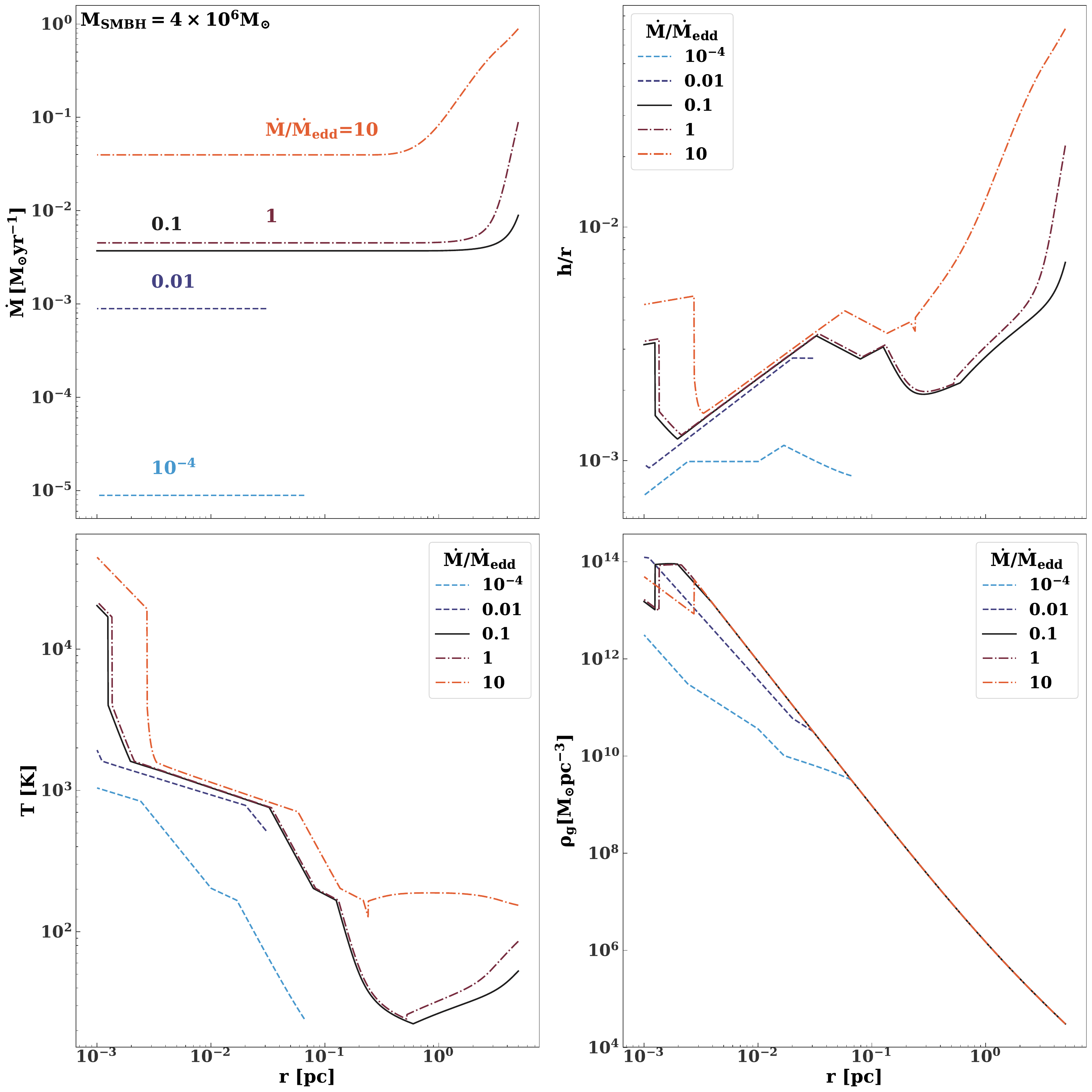}
    \caption{(Clockwise from top left) Radial profiles of mass accretion rate, aspect ratio, gas temperature, and gas density. Different colors correspond to different mass accretion rates at the outer boundary. The equation describing disk structure are in \citet{thompson+2005}.}
    \label{fig:tprofiles}
\end{figure*}

We consider an MBH mass of $4\times 10^6 \  M_{\odot}$ like Sgr A$^{\star}$. Unless otherwise specified, we adopt the \citet{thompson+2005} AGN disk model, with the modifications described in \citealt{Tagawa2020}. 
Figure~\ref{fig:tprofiles} shows radial profiles of gas density, temperature, and scale height for different mass accretion rates onto the MBH. Outside of the central $\sim 1$ pc the disk is Toomre unstable and forms stars. Heating from radiation pressure and supernovae maintains the disk in a marginally stable state. The density profile is set by marginal Toomre stability and is independent of the mass accretion rate. The gas density on large scales is $\rho_g\approx 10^6 \left(r/{\rm 1 pc}\right)^{-3} M_{\odot}$ pc$^{-3}$. For the fiducial accretion rate in \citet{Tagawa2020} ($0.1 \dot{M}_{\rm Edd}$) the gas temperature is $20$ K at $\sim 1$ pc. (Corresponding to a sound speed of $\sim 0.4$ km s$^{-1}$. The disk is very thin with aspect ratio, $h/r \approx 10^{-3}-10^{-2}$). We also consider lower accretion rates down to $10^{-4} \dot{M}_{\rm Edd}$, where there is no self-consistent solution extending to pc scales. In such cases we assume an $\alpha$-disc, whose outer radius is set by Toomre instability. 

We now estimate the capture rates of different types of binaries within the disc: BH-BH, BH-star, and star-star. 
We assume a (number) density profile for the BHs of  
\begin{align}
    &n_{\rm bh}(r)=n_o \left(\frac{r}{r_o}\right)^{-2} \left(\frac{h}{r}\right)^{-1}\nonumber\\
    &n_o=\frac{N_{\rm bh}}{4 \pi r_o^3},
\end{align}
where $N_{\rm bh}$ is 1000 and $r_o$ is the outer radius of the BH distribution ($3 \rm{pc}$). This is similar to the initial number of disk black holes in \citet{Tagawa2020}, where the black hole component is flattened and rotating. This was suggested to occur via vector resonant relaxation \citep{szoelgyen&kocsis2018}. However, the degree of flattening and hence the number of disk black holes will depend on the mass function, and for realistic conditions, it is not clear whether indeed such a flattened disk of BHs should indeed exist, nor why should it be alligned with the gaseous AGN disk. Nevertheless, in order to compare with the results of \citet{Tagawa2020}, we consider similar conditions. Alternatively, multiple star-formation epochs might give rise to new generation of BHs that form in the AGN disk \citep{Stone2017}, and provide a large number of BHs in the disk. The velocity distribution of the BHs is taken as a Maxwellian whose scale parameter is $(h/r) v_{\rm kep}$, where  $v_{\rm kep}$ is Keplerian velocity. For simplicity we assume all BHs are 10 $M_{\odot}$.

\begin{figure}
    \includegraphics[width=\columnwidth]{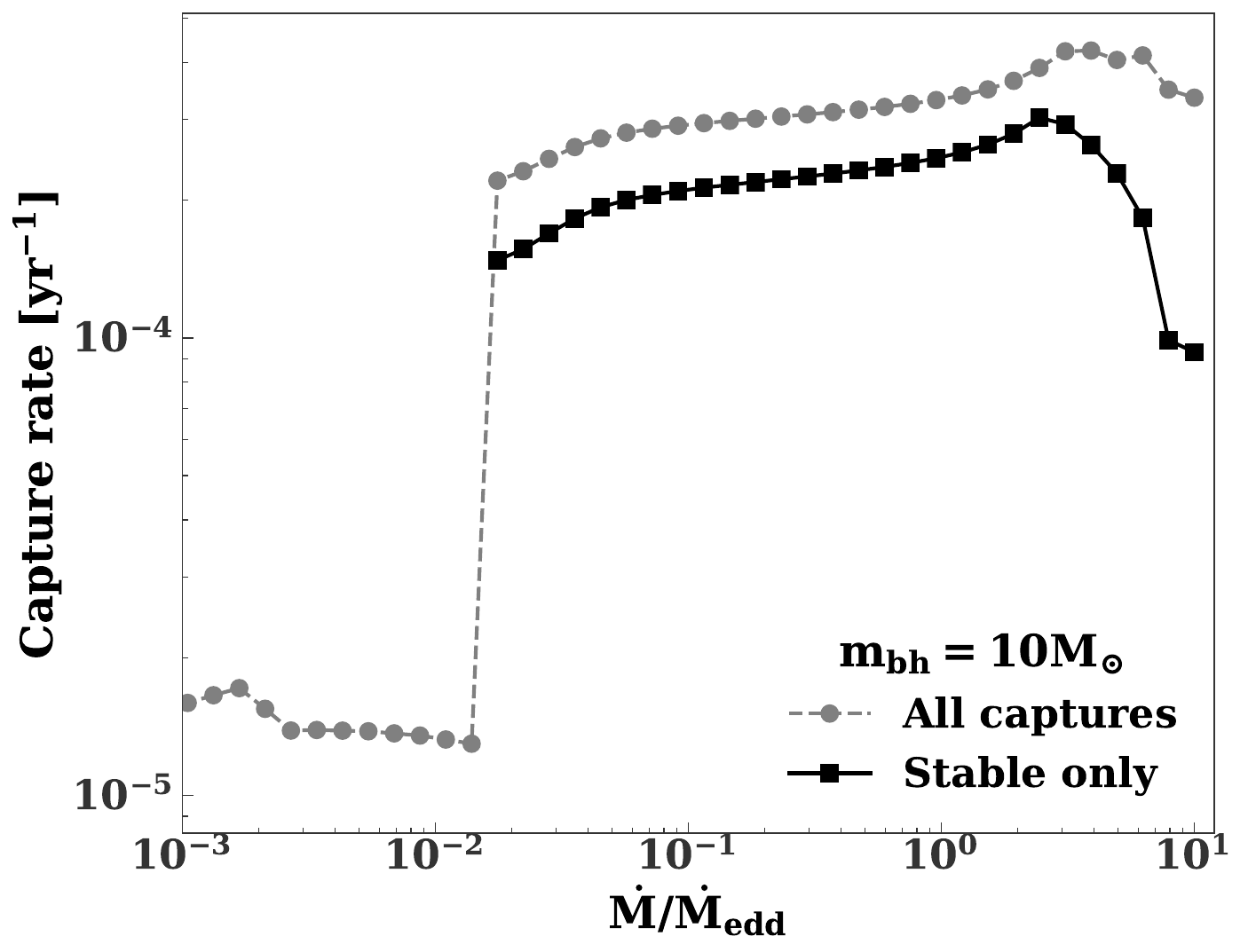}
    \caption{Total BH-BH capture rate due to GDF in model accretion disks (see text and Figure~\ref{fig:tprofiles} for details). Both black holes are $10 M_{\odot}$. The dashed, gray line shows the total capture rate, while the solid, black line shows the rate of captures in the unfocused regime. Only the latter will form stable, long-lived binaries (see the discussion in \S~\ref{sec:focusStability}).}
    \label{fig:captRateAgn}
\end{figure}

Figure~\ref{fig:captRateAgn} shows the BH-BH capture rate  from equation~\eqref{eq:captureRate} integrated over the entire disk as a function of mass accretion rate (at the outer boundary), viz.
\begin{equation}
 \Gamma_{\rm tot}= \int_{r_{\rm min}}^{3 {\rm pc}} n_{\rm bh}(r) \Gamma(r) 4 \pi r h(r) dr.  
\end{equation}
The total rate is dominated by large scales and is a weak function of $r_{\rm min}$.
The dashed line shows the total capture rate, while the solid line shows the rate of captures in the unfocused regime. Considering the discussion in \S~\ref{sec:focusStability}, only the latter can lead to long-lived, stable binaries. Thus we expect a total binary formation rate of a few $\times 10^{-4}$ per year in disks with Eddington ratios $\gtrsim0.01$. This is comparable to the formation rate calculated by \citet{Tagawa2020} (cf their Figure 7). The average capture rate per black hole is $2\times 10^{-7}$ per year. 

\begin{figure}
    \includegraphics[width=\columnwidth]{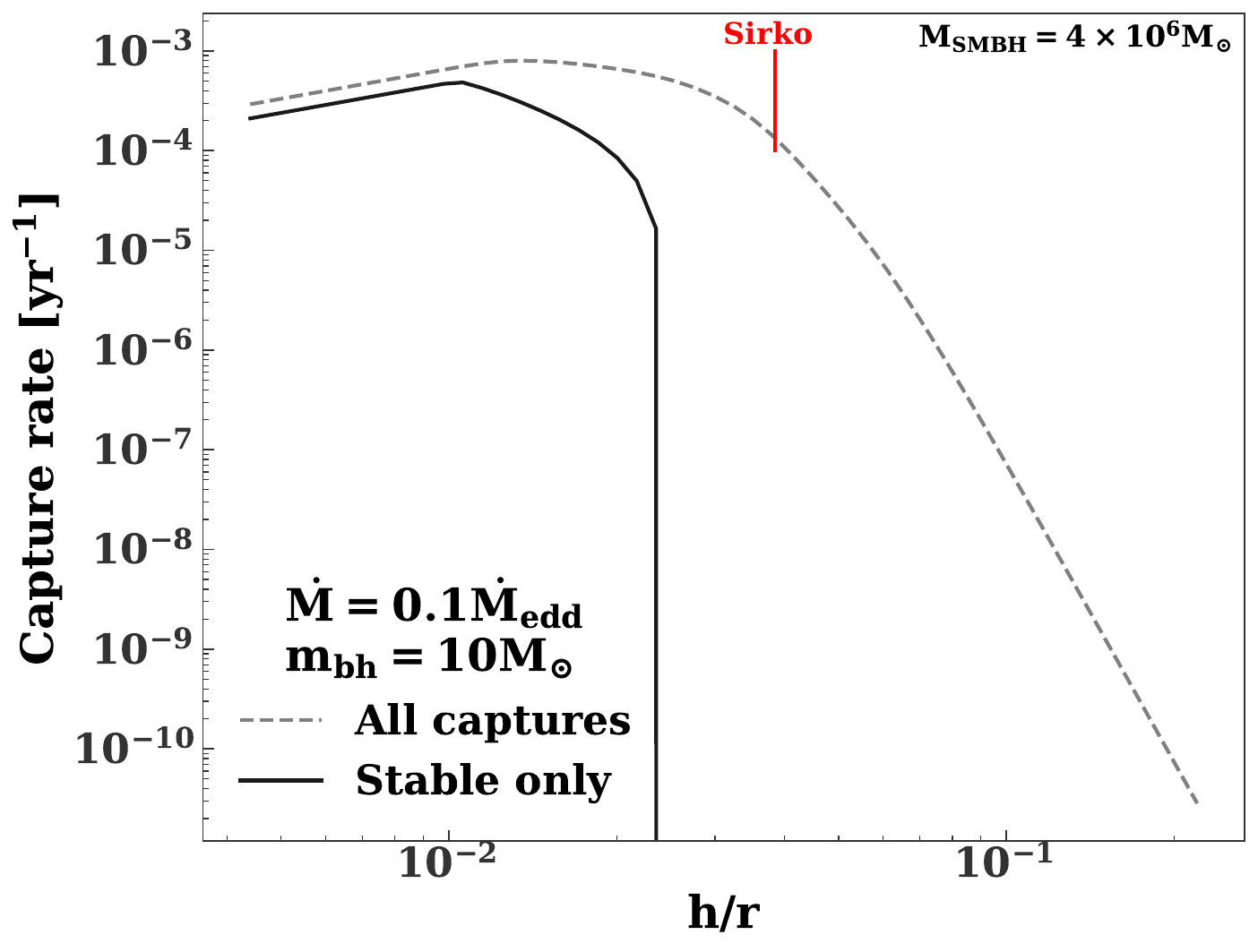}
    \caption{Total BH-BH capture rate for artificially inflated AGN disks with $\dot{M}/{M}_{\rm Edd}=0.1$ (i.e. we use the \citealt{Tagawa2020} model in Figure~\ref{fig:tprofiles}, but aspect ratio is rescaled by a constant factor). The x-axis shows the aspect ratio at 3 pc after rescaling. If the disk aspect ratio is increased by an order of magnitude, the rate of stable captures goes to 0 (see the discussion in \S~\ref{sec:focusStability}). The vertical red line shows the aspect ratio for the \citet{sirko&goodman2003} model for fiducial parameters 
    \label{fig:aspectRatioEffect}}
\end{figure}

However, this result is sensitive to the aspect ratio of the disc, as shown in Figure~\ref{fig:aspectRatioEffect}. This figure shows the BH-BH capture rate after artificially rescaling the aspect ratio of the $\dot{M}/\dot{M}_{\rm edd}=0.1$ disk in Figure~\ref{fig:tprofiles}. For very thin disks, the capture rate increases linearly with the velocity dispersion and with the aspect ratio. However, for thicker disks the capture rate falls off steeply with aspect ratio, because captures only come from the tail of the velocity distribution. The capture rate of stable binaries is 0 for aspect ratios above a few$\times 10^{-2}$.

The aspect ratio depends on the mechanism for angular momentum transport. In the above calculations, the radial gas velocity is 0.15 times the local sound speed in the outer disc, as in \citet{Tagawa2020}. In \citet{sirko&goodman2003}, the radial velocity is $\approx \alpha (h/r)$ times the sound speed and can be much smaller.
For $\alpha=0.1$ and $\dot{M}/\dot{M}_{\rm Edd}=0.1$, the aspect ratio at parsec scales is approximately an order of magnitude larger than in the \citet{Tagawa2020} model. Thus, the total capture rate is a factor of $\sim 2$ smaller, and the rate of {\emph stable} captures is 0.

So far we have focused on BH-BH captures. However BH-star captures and star-star captures will also occur.  We expect a few $\times 10^6$ stars old, low mass ($\lsim M_{\odot}$) stars within the central $\sim 3$ pc of the Galaxy. Geometrically, we expect $\sim 10^4$ stars within the disc. We estimate the rate of BH-star captures to be $\sim 2\times 10^{-3}$ yr$^{-1}$ and the rate of star-star captures to be $\sim 10^{-2}$ yr$^{-1}$.\footnote{For simplicity we assume all stars are $1 M_{\odot}$.} This assumes the low mass stars in the disk have an $r^{-2}$ density profile like the black holes. However studies of relaxation in spherical clusters, the density profile of low mass species falls between $r^{-1.5}$ and $r^{-1.75}$ \citep{alexander&hopman2009}. For an $r^{-1.5}$ stellar density profile the BH-star and star-star capture rates are $\sim 1.4 \times 10^{-3}$ and $6\times 10^{-3}$ yr$^{-1}$, respectively.
Thus, the overall capture rate per object is $\sim$1 per Myr. Note that the capture rate is dominated by stellar captures. For example, the BH-star capture rate is roughly one order of magnitude greater than the BH-BH capture rate.

\subsection{Gas-enriched globular/massive clusters}\label{subsub:GC}
 For decades, globular cluster (GCs) were thought to contain a single-aged stellar population, i.e. originating from a single burst of star formation. However, over the last two decades it was found that the vast majority of GCs host at least two
 or even more population of stars (see detailed reviews in \citealp{Renzini2015,Bastian2018,Gratton2019}), which were suggested to form at different epochs.
Although the exact origin of the multiple populations is still unknown, their existence is a smoking gun for gas-replenishment in GCs. As we pointed out in \cite{RoznerPerets2022}, the dynamics and evolution of binaries in GCs should be revised, and the gas involved in the formation of the second or further generation should affect the evolution of previously-formed stellar populations and binaries which become embedded in such gas-rich environment \cite{Maccarone2012,Leigh2013,Leigh2014,Roupas2019,RoznerPerets2022}. Similar to AGN disks, we therefore might expect that high gas abundance could also potentially give rise to gas-assisted binary formation in these environments. 

The density of gas originating in the epoch of second (or further) star formation is highly uncertain (e.g. \citealp{Bastian2018}) and can be roughly estimated by $\rho_g\sim M_{2P}/V_{2P}\sim 10^5 \ M_\odot \ \rm{pc}^{-3} $, where $M_{2P}$ is the mass of the second population and $V_{2P}$ is the volume in which it is enclosed, if we again consider specifically BHs, the typical BH number density is $n_\bullet\sim 10^3 \ \rm{pc}^{-3}$. The temperature during the star forming stage may differ from the current temperature. Hence, following \cite{Bekki2010}, we consider a gas temperature of $100 \ K$, which corresponds to a sound speed of $0.6 \ \rm{km/sec}$. Following \citet{Bekki2010}, \citet{MastrobuonoBattistiPerets2013}, and \citet{MastrobuonoBattisiPerets2016}, 
we consider the second population of stars as embedded in a disk, with an aspect ratio of $h/r\sim c_s/v_K\sim 3\times 10^{-2}$, and a radius of $1 \ \rm{pc}$. Unless stated otherwise, these will be the fiducial parameters. We assume a stellar density of $n_\star=10^4 \ \rm{cm^{-3}}$ for Solar star mass. The mass density of each stellar species is constant i.e. $\rho \equiv \rm{constant}=m_i n_i$, where $\rho$ is a constant. 

The total mass of the cluster is $2\times 10^5 \ M_\odot$.

\begin{figure*}
    \includegraphics[width=\columnwidth]
    {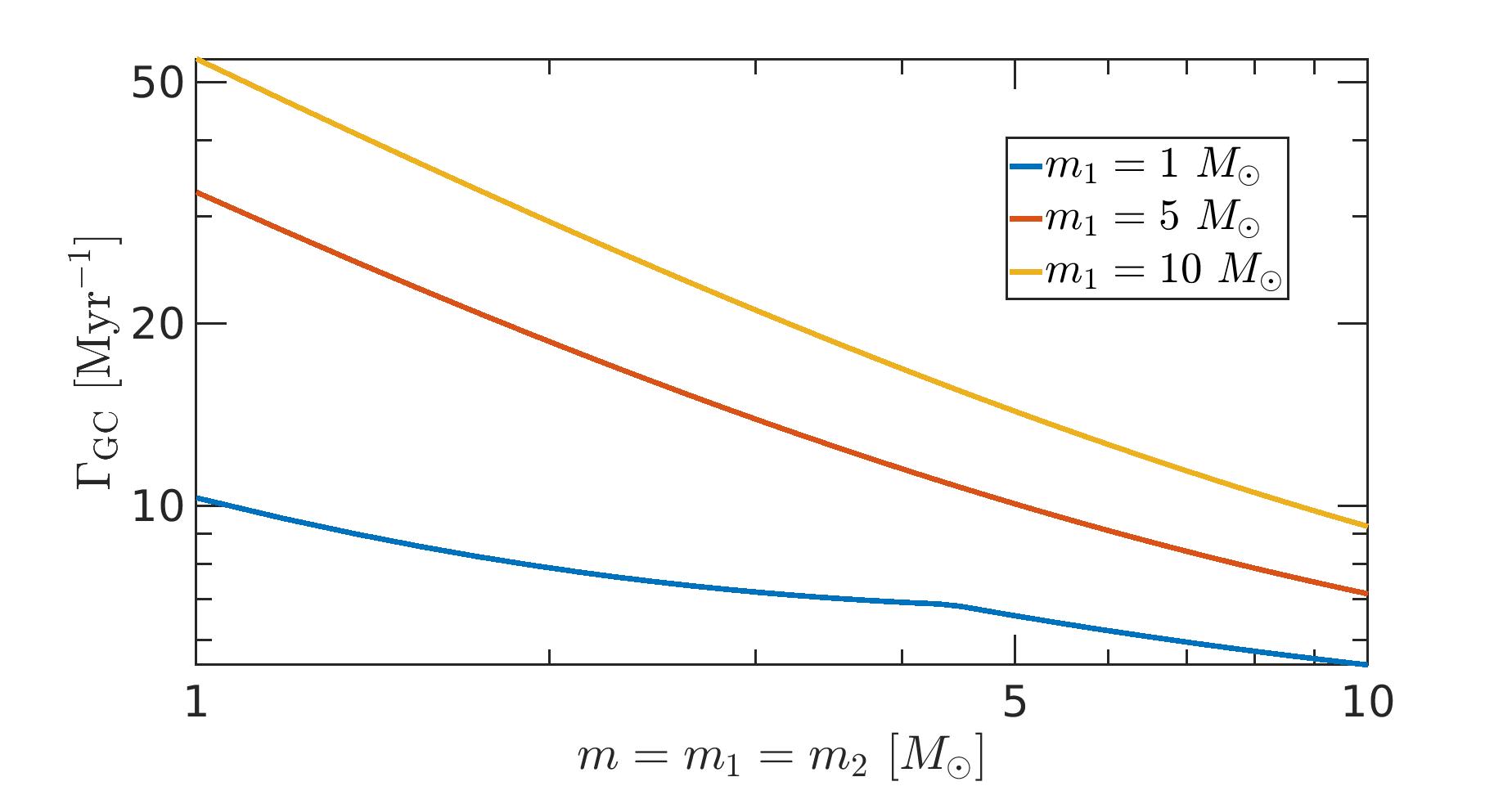}
    \includegraphics[width=\columnwidth]
    {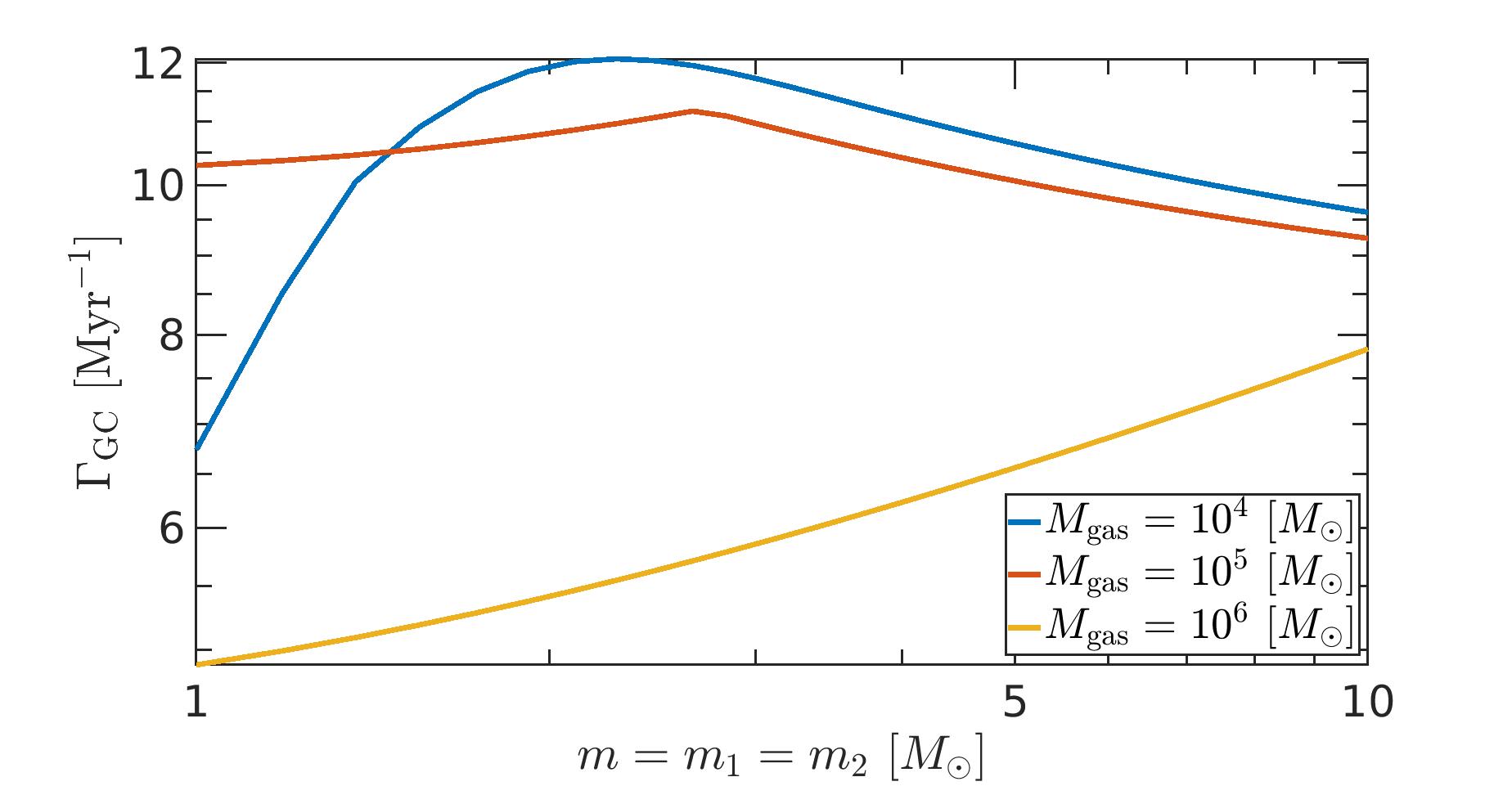}
    \caption{Left: The capture rate per object for different masses, in a second (or later) star generation environment in a GC, for our fiducial model specified above. Right: The capture rates for equal masses, given different gas masses.}
    \label{fig:GC}
\end{figure*}

In Figure~\ref{fig:GC}, we present the capture rates per object in multiple-population gas-enriched environment in GCs (or equivalent younger massive clusters objected in other galaxies), or nuclear clusters which do not host MBHs. As can be seen on the left panel, the capture rate decreases for larger masses of captured objects, although the total capture rate is high for all the mass range. The right panel shows the capture rate for different gas and stellar masses. As can be seen, the dependence on the gas mass is not trivial, as it also modifies the Hill radius, which depends on the enclosed stellar and gaseous mass. Sufficiently high densities lead to smaller probabilities for capture. 
 The overall dependence on the central mass is stronger than the dependence on the gas density. 

As we showed in \cite{RoznerPerets2022}, the merger of pre-existing binaries in such environments could be catalyzed by the gaseous environment; the additional gas-catalyzed formation binaries would therefore also further increase the binary mergers rate. We conclude that a significant fraction of the merged objects in this environments are the product of gas-assisted mergers, similarly to the conclusion in \cite{Tagawa2020} for AGN disks.  That being said, this conclusion, like our results on AGN disks, strongly depend on the existence of a relatively flattened disk; thicker disks would not allow for significant capture rates.

\subsection{Protoplanetary disks}\label{subsub:PPD}

The early stages of planet formation take place in protoplanetary disks that initially contain gas, with typical dust-to-gas ratio of $\sim 1\%$ \citep{ChiangGoldreich1997}. Planetesimals vary in size, and the nature of their interaction with gas changes accordingly \citep{Weidenschilling77}. While small particles are well-coupled to the gas, large planetesimals interact differently and their interaction with the gas could be modeled using GDF \citep{Gri+15,GrishinPerets2016}. \cite{GoldreichSari2002} suggested that two initially unbound objects in the Kuiper-belt could form a wide binary with comparable masses, via dissipation induced by dynamical friction. Due to the high abundance of gas on the early stages of planet formation, {\emph gas} dynamical friction could potentially play a similar role and lead to gas-assisted binary formation in protoplanetary disks. 

Following \cite{Armitage2010,PeretsMurrayClay2011}, we consider a background gas density of $\rho_g = 3\times 10^{-9}(a/\rm{AU})^{-16/7} \ \rm{g \ cm^{-3}}$ and sound speed of  sound speed of $c_s=0.6 (a/AU)^{-3/8} \ \rm{km/sec}$. Following \cite{GoldreichSari2002} we choose a separation of $a=40 \ \rm{AU}$ from a Solar mass star and for a $100 \ \rm{km}$ objects, the typical surface density is $\Sigma\sim 3\times 10^{-4} \ \rm{g \ cm^{-2}}$. Then, the typical corresponding density is $\rho \sim \Sigma/r$ where $r$ is the distance from the Sun, such that the number density of candidates for capture in this mass is $n_b= \rho/m$
. Given these assumptions, a capture of $\sim 10^{22} \ \rm{g}$ object is not likely to occur via GDF within the gas lifetime. Since the typical velocity dispersion of large objects relative to the gas is supersonic, the energy dissipation induced by gas-dynamical friction it not efficient enough to enable such a capture, such that dynamical friction induced by smaller bodies will be more efficient under these assumptions.

\section{Potential caveats}\label{sec: caveats}
Here we will briefly discuss potential caveats of our model:
\begin{enumerate}
\item Interaction of objects in gaseous environments could be affected not only by GDF, but by migration in circumbinary disks (e.g. \citealp{McKernan2012,Stone2017}), or Bondi-Hoyle accretion \citep{AntoniMacLeodRamirezRuiz2019}. Later stages of the evolution are more likely to be dominated by Bondi-Hoyle accretion rather than GDF. A more detailed description of the gaseous interaction is beyond of the scope of this paper and is left out for future studies. 

\item We showed that headwind will have a minor effect on capture (see \S~\ref{subsec:headwind}). However, we considered only a linear constant headwind, and neglected shear.

\item We use \citet{Ostriker1999}'s prescription for GDF, which assumes straight-line trajectories. However, objects will have curved orbits following capture. This prescription also neglects interference between each object's wake. Nevertheless, while these could be critical for the evolution and inspiral of bound binaries, the conditions for the initial capture are generally consistent with our assumptions. Future hydrodynamical simulations may help resolve the potential importance of this issue.

\item Gas accretion is not taken into account here, and could significantly change the mass distribution of objects in gaseous environments, as well as the heating rates. However, the dynamical timescale for the capture is relatively short, and we do not expect mass-gain to be of significant importance for the capture. Nevertheless, accretion feedback could potentially change the GDF effect (see \citealt{GruzinovNDF+2020} and the references therein). 

\item Close to the Hill radius, the interaction of the two stars with the external potential could give rise to temporary captures, which potentially allow for longer close interaction and more significant energy dissipation, and hence higher capture rates (see \citealp{PetitHenon1986,BoekholyKocsis2022} and references therein). It should be noted again that in this paper we focus on the capture process only and leave out further evolution, including stability for future studies (in prep). Such evolution will be affected also by the external potential as well as other dissipation mechanisms such as GW radiation (e.g. \citealp{LiLaiRodet22,BoekholyKocsis2022})
\item Feedback effects from the energy deposited into the gas due to the capture and subsequent migration could potentially change the conditions of the gaseous environment. We consider this issue in the following section.

\end{enumerate}

\section{Feedback: heating and cooling}\label{subsec:heating}

Up to this point, we have considered the effect of the gas on the capture-formation of binaries. However, the capture, and more importantly the later inspiral of the binaries in gas could potentially give rise to feedback and heat the gaseous environments, potentially quenching further gas-assisted formation of binaries. In the following we show that the heat generated due to the gas dissipation is mostly radiatively emitted and does not contribute significantly to the disk heating. Other processes such as gas accretion onto stars/compact-objects might provide additional feedback (e.g. through jets); modeling the effect of these process depends on many different assumptions and uncertainties, and is beyond the scope of the current study. 

Capture and inspiral of binaries in the gas could heat the environment. The heating energy $E_{\rm heat}$ from the inspiral could be approximated by 

\begin{align}
E_{\rm heat}\approx  \frac{Gm_1m_2}{2a_{\rm final} },
\end{align}
where $a_{\rm final}$ is the final semimajor axis of the binary prior to merger.
Then, the average heating rate per area is

\begin{equation}
    \ell_{\rm heat}=E_{\rm heat} \gamma,
\end{equation}

\noindent 
where $\gamma$ is the binary capture rate per area, i.e. $\gamma=\Gamma_{\rm cap}/A$  where $\Gamma_{\rm cap}$ is the capture rate as we calculated earlier, and $A$ is a typical area.
The typical cooling rate could be approximated by blackbody cooling,

\begin{align}
\ell_{\rm cool}= \sigma_{\rm SB}T_{\rm eff}^4
\end{align}
Here we compare between the cooling and heating rates for the different environments we discussed, for our fiducial models specified in Table \ref{table:capture rates}.

\subsubsection{SF environments}

The typical Hill radius is $R_{\rm Hill}=2\times 10^4 \ \rm{AU}$, and the cloud lifetime is $5 \ \rm{Myr}$ (\citealp{BerginTafalla2007} and references therein). These parameters yield

\begin{align}
\ell_{\rm cool}\approx 5\times 10^{36} \left(\frac{T_{\rm eff}}{10 \ \rm{K}}\right)^4 \ 
\rm{erg \ sec^{-1} \ pc^{-2}}
\end{align}
From integration of $da/dt$ (as derived in \citealt{RoznerPerets2022}) for $5 \ \rm{Myr}$, and taking the relative velocity between the gas and the objects as half the Keplerian velocity, the separation after $5 \ \rm{Myr}$, $a_{\rm final}$ is $ \approx 380 \ \rm{AU}$ , hence

\small
\begin{align}
\ell_{\rm heat}= 
1.68 \times 10^{34}
\left(\frac{m}{1 \ M_\odot}\right)^2\left(\frac{380 \ \rm{AU}}{a_{\rm final}}\right)\left(\frac{\gamma}{1.4\rm{Myr^{-1} pc^{-2}}}\right) \ \rm{erg \ sec^{-1} pc^{-2}}
\end{align}
\normalsize

\noindent
We then conclude that cooling is efficient in SF environments.

\subsubsection{Second generation in GCs}

For second generation gas embedded in a disk, the energy radiated away within gas lifetime of $50 \ \rm{Myr}$ is given by 

\begin{align}
\ell_{\rm cool}\approx 5.1\times 10^{40} \ \rm{erg \ sec^{-1} \ pc^{-2}}
\end{align}

\noindent
and the heating energy

\small
\begin{align}
\ell_{\rm heat}= 2.92\times 10^{36} \left(\frac{m}{10 \ M_\odot}\right)^2 \left(\frac{0.07 \ \rm{AU}}{a_{\rm final}}\right)\left(\frac{\gamma}{\rm{Myr^{-1} \ pc^{-2}}}\right) \ \rm{erg \ sec^{-1} \ pc^{-2}}
\end{align}
\normalsize
Hence, cooling is efficient also here.

\subsubsection{AGN disks}
 
Over the $10^7-10^8 \ \rm{yr}$ lifetime of an AGN disk a $10 M_{\odot}$ BBH (binary black hole) at $1 \ \rm{pc}$ can inspiral to $\sim 1 \ \rm{AU}$. At $1 \ \rm{pc}$
\begin{equation}
    \ell_{\rm heat}=4\times 10^{35}  \left(\frac{m}{10 \ M_\odot}\right)^2 \left(\frac{1 \ \rm{AU}}{a_{\rm final}}\right) {\rm erg\,\, s^{-1}\,\, pc^{-2}}
\end{equation}
for the $\dot{M}/\dot{M}_{\rm edd}=0.1$ model. On the other hand, the cooling luminosity per unit area at 1 pc is
\begin{equation}
    \ell_{\rm cool}=\sigma_{\rm sb} T_{\rm eff}^4\approx 5\times 10^{37} {\rm erg\,\, s^{-1}\,\, pc^{-2}}
\end{equation}
Thus, we do not expect black hole-black hole captures to significantly perturb the disc. The heating contribution from stellar captures is smaller.

Thus, we do not expect significant heating of the gas via binary inspiral, as cooling dominates heating in all environments we consider.
Moreover, the heating rate may be overestimated. Binaries can merge by eccentricity excitation before significantly inspiraling in semimajor axis (see Figure~\ref{fig:examples}). 

\section{Discussion and summary}\label{sec:summary}
The evolution of binaries in gaseous environments was extensively studied over the last few years, in the context of various physical environments, and in particular AGN disks. Here we focused on binary {\emph formation} rather than the later stages of the evolution of pre-existing binaries. We made use of analytic arguments also validated with few-body simulations to derive the criteria for gas-assisted binary capture in different astrophysical environments, and discussed its expected rates and implications. We showed that interaction with gas could play a key role in binary formation, depending on the specific conditions of the gaseous and overall environment. We also pointed out several potential caveats, and potential processes that may affect these issues, but are not considered in depth in the current study. The conditions we derived are general and could be applied in principle to any type of gas-rich environment and used to characterize the formed gas-assisted binary population. 

Here we considered several typical gas-rich environments and conditions where gas-assisted binary formation could occur, including star-forming regions, AGN disks, gas-enriched clusters and young protoplanetary disks. We find that all of these environments, besides protoplantary disks, support high rates of gas-assisted binary formation, and thereby this formation channel is expected to significantly affect the binary population and its properties in these environments. 
In the following we briefly discuss the implications of gas-assisted captures in specific environments.

\subsection{Implications for different environments}
\begin{itemize}
    \item Star-forming regions: Gas-assisted binary formation could then prove to be a major channel for the general formation of stellar binaries in star-forming regions, and hence in the universe at large. 
    
\item Gas-enriched clusters: In gas-enriched globular clusters, gas-assisted binary formation could alter the binary population during the early hundred Myrs of evolution, if such clusters were gas enriched, as suggested by the existence of multiple stellar populations. Currently used models of globular cluster stellar populations and their evolution do not consider such gas phase, nor its implications for the binary population and evolution. Fundamental aspects of such models should therefore be potentially reconsidered. In addition, the gaseous environment may also give rise to high productions rates of GW sources, even higher than those found by us in \cite{RoznerPerets2022}, where we focused only on {\emph primordial} binaries, where capture-formed binaries could further increase the rates making this a potential key channel for the origin of GW sources from stellar compact-object binaries. 

\item AGN disks: For the case of AGN disks, we pointed out that the required conditions for efficient capture involve a very thin disk. Although such conditions might exist close to the MBH, it is not clear that such large-scale thin disks exist, and observations of large-scale maser disks suggest such disks are in-fact very thick \cite[e.g.][and references therein]{Yam+04,May+09}. Studies suggesting high production rates of gravitational-waves sources in AGN environments rely on a high supply rate of BH binaries into the disk. Since we find that such high supply rates can only likely be accommodated by the existence of large-scale (pc scale) disks, and these might be short-lived or rare (if they exist at all), we suggest considering the AGN channel for GW sources with caution. We do note that the existence of a young stellar disk in the Galactic center
\citep{LevinBeloborodov2003}
suggest the past existence of at least a short-lived large-scale thin gaseous disk (it had to be thin to allow for star-formation), but this case is quite different than that envisioned for AGNs.

\item Protoplanetary disks: Dynamical friction assisted binary formation was first suggested in this context by \cite{GoldreichSari2002}, and it was shown to be highly efficient. Here we find that the gas-phase and the generalization to gas dynamical friction does not give rise to higher rates, as discussed above, and therefore play a lesser role in binary planetesimal formation in such environment.  
\end{itemize}

Finally, we point out that sequential multicaptures may occur and give rise to fast growth of objects, and/or to the formation of high multiplicity systems which later become unstable.
Such multicaptures are expected to take place whenever more than one capture occurs per gas lifetime.
Detailed study of multicaptures is beyond the scope of the current study but will be explored in depth in a dedicate study (in prep).

\section*{Acknowledgments}

We thank the referee, for his helpful and constructive comments. MR acknowledges the generous support of Azrieli fellowship. AG is supported at the Technion by a Zuckerman Fellowship.
We thank Hiromichi Tagawa for insightful conversations.

\section*{Data Availability}

The data that support the findings of this study are available from the
corresponding author upon reasonable request. 


\bibliographystyle{mnras}






\appendix
\section{Analytic solution for velocities}
\label{sec:vel}

Here we present approximate closed-form solution for the velocity as a function of time. 
If $I(v/c_s)\approx\ln \Lambda$, as in the supersonic regime, then

\begin{align}
    &v_{\rm sup}(t)=v_i \left(1-\frac{t}{t_d}\right)^{1/3}, \nonumber\\
    & t_d=\frac{v_i^3}{12 \pi G^2 m \rho_g \ln \Lambda}
\end{align}
The velocity decays to zero at $t_d$, but before this happens, the velocity will become subsonic. In this case $I\approx \frac{v^3}{3 c_s^3}$, and
\begin{align}
    &v_{\rm sub}(t)=v_o \exp{\left[\frac{-(t-t_i)}{\tau}\right]},\nonumber\
    \tau=\frac{3 c_s^3}{4 \pi G^2 m \rho_g}
\end{align}

With the above solutions we can compare the length scales over which the stars in the binary decelerate ($\ell_d$). In the supersonic case, 
\begin{align}
\frac{\ell_{d, 1}}{\ell_{d,2}}&=\frac{v_{i,1} t_{d,1}}{v_{i,2} t_{d,2}} =\frac{v_{i,1}^4}{v_{i,2}^4}\frac{m_2}{m_1}=q^5.
\end{align}
Above, subscripts 1 and 2 denote the primary and secondary star of the binary respectively, and $q$ is the ratio between the secondary and primary mass. In the subsonic case, 
\begin{align}
\frac{\ell_{d, 1}}{\ell_{d,2}}&=\frac{v_{i,1} \tau_{1}}{v_{i,2} \tau_{2}} 
        =\frac{v_{i,1}}{v_{i,2}}\frac{m_2}{m_1}=q^2.
\end{align}

\section{Detailed derivation of the threshold velocities for capture}
\label{sec:Threshold Derivation}

Here we outline the derivations of the threshold velocities in Table~\ref{table:Critical velocities, vg=0} in more detail case by case
\subsection{Supersonic, unfocused threshold}
In the supersonic, unfocused limit the work done on star i is
\begin{equation}
W_i\approx \mathbf{F_{GDF, i} \cdot \ell_i} \approx \frac{4 \pi G^2 \ln(\Lambda) \rho_g m_i^2}{v_i^2} \ell_i,
\end{equation}
where $m_i$ is the mass, $v_i$ is the velocity, $\rho_g$ is the gas density, $\ln \Lambda$ is the 
Coulomb logarithm and $\ell_i$ is the path length traveled. We assume the binary center-of-mass is at rest with respect to the gas. Thus,
the masses, velocities, and path lengths can be rewritten in terms of the total mass ($m_{\rm bin}$),
the mass ratio ($q$), and the Hill radius ($R_{\rm Hill}$), viz.
\begin{align}
&m_1=\frac{m_{\rm bin}  }{1+q}, \
m_2=\frac{m_{\rm bin} q}{1+q}\nonumber\\
&v_1=\frac{v_{\infty}  q}{1+q}, \
v_2=\frac{v_{\infty}   }{1+q}\nonumber\\
&\ell_1 \approx q^5 R_{\rm Hill} \sqrt{1-\frac{b^2}{R_{\rm Hill}^2}}, \
\ell_2 \approx R_{\rm Hill} \sqrt{1-\frac{b^2}{R_{\rm Hill}^2}}.
\end{align}
 Above $b$ is the impact parameter. We use the initial velocity to estimate the energy dissipated. This is justified because most energy will be dissipated at large velocities, due to the quadratic dependence of kinetic energy on velocity. The path length of the secondary star is simply the straight-line distance through the Hill sphere.\footnote{More precisely $\ell_2 = \frac{2 R_{\rm Hill}}{1+q} \sqrt{1-\frac{b^2}{R_{\rm Hill}^2}}$} The path length of the primary is $q^5$ times this distance from  the preceding Appendix. Then, the total work 
 done is 
\begin{equation}
W_{\rm tot } =\frac{4 \pi G^2 m_{\rm bin}^2 \ln(\Lambda) \rho_g R_{\rm Hill} q^2 (1+q)}{v_{\infty}^2}\sqrt{1-\left(\frac{b}{R_{\rm Hill}}\right)^2}
\end{equation}

Finally, we equate $W_{\rm tot}$ with the energy of the unbound orbit $\left(\frac{1}{2} \frac{m_{\rm bin} q}{(1+q)^2} v_\infty^2\right)$ to obtain the threshold capture velocity 
\begin{equation}
v_c=\underbrace{\left(8 \pi G^2 \rho_g m_{\rm bin} R_{\rm Hill} \ln \Lambda \right)^{1/4}}_{v_x} q^{1/4} \left(1+q\right)^{3/4} \left(1-\frac{b^2}{R_{\rm hill}^2}\right)^{1/8},
\label{eq:vx}
\end{equation}
For simplicity, we drop the last term.
\subsection{Subsonic, unfocused threshold}
The derivation for the subsonic, unfocused case is similar, except the work done on star i is 
\begin{equation}
W_i\approx \frac{4 \pi G^2 \ln(\Lambda) \rho_g m_i^2}{3 c_s^3} v_i \ell_i,
\end{equation}
and the path length of the primary star is $q^2$ times the path length of the secondary (see Appendix~\ref{sec:vel}). Then the velocity threshold is 
\begin{equation}
v_{c}=\underbrace{\frac{8 \pi G^2 \rho_g m_{\rm bin} R_{\rm Hill}}{3 c_s^3}}_{v_s} q \sqrt{1-\frac{b^2}{R_{\rm Hill}^2}}
\end{equation}

\subsection{Supersonic, focused threshold}
In the focused regime, we approximate the stellar trajectories as parabolic. The separation between the stars, $r$, and their relative velocity $v_{\rm rel}$, are 
\begin{align}
&r=\frac{2 x R_{\rm Hill}}{1+\cos(f)}\nonumber\\
&v_{\rm rel}=\sqrt{\frac{2 G m_{\rm bin}}{r}},
\end{align}
where $f$ is the true anomaly of the oribt and $x$ is the pericentre distance in units of the Hill radius. In the supersonic regime, the work done on star i is 
\begin{align}
W_i&=4 \pi G^2 \rho_g m_i^2 \ln (\Lambda) \int_{t_{\rm start}}^{t_{\rm end}} v_i^{-1} dt = \nonumber\\
&=4 \pi G^2 \rho_g m_i^2 \ln (\Lambda) \int_{-f_{\rm end}}^{f_{\rm end}} v_i^{-1} \sqrt{\frac{(x R_{\rm Hill})^3}{2 G m_{\rm bin}}} \sec(f/2)^4 df.
\end{align}
Above $\pm f_{\rm end}=\pm \arccos (2 x -1)$ are the true anomalies where the distance between the stars exceeds the Hill radius. Thus, the critical velocity capture 
\begin{equation}
v_c=\frac{v_x^2}{v_{\rm esc}} \frac{\sqrt{1+q+q^3+q^4}}{q} h(x),
\end{equation}
where $v_x$ is defined in equation~\eqref{eq:vx} and $v_{\rm esc}$ is the escape velocity at the Hill radius.
$h(x)$ is a complicated function. However, $1<h(x)<2$, except very close to $x=1$, and is dropped for simplicity. We also drop the last two terms under the square root. (Note $0<q\leq 1$.)

\subsection{Subsonic, focused threshold}
In the subsonic limit, the work done on star i is
\begin{align}
W_i&=\frac{4 \pi G^2 \rho_g m_i^2}{3 c_s^3} \int_{t_{\rm start}}^{t_{\rm end}} v_i^{2} dt = \nonumber\\
    &=\frac{4 \pi G^2 \rho_g m_i^2}{3 c_s^3} \int_{-f_{\rm end}}^{f_{\rm end}} v_i^{2} \sqrt{\frac{(x R_{\rm Hill})^3}{2 G m_{\rm bin}}} \sec(f/2)^4 df.
\end{align}

Thus, the velocity threshold is 
\begin{equation}
v_c=\sqrt{8 v_{\rm esc} v_s} \frac{\sqrt{q}}{1+q} (1-x)^{1/4}.
\end{equation}

For simplicity, we drop the last term on the right-hand side. 

Above we have ignored some edge cases. For example, the secondary may be supersonic, while the primary is subsonic. However, the approximate thresholds derived here are within a factor of $\sim$2 of the thresholds from our few-body simulations and are adequate for our purposes.

\bsp	
\label{lastpage}
\end{document}